\documentclass[]{spie}  

\usepackage{amsmath,amsfonts,amssymb}
\usepackage{graphicx}
\usepackage[colorlinks=true, allcolors=blue]{hyperref}
\usepackage{pstricks}
\usepackage{cancel}
\usepackage{multirow}
\title{General formalism for Fourier based Wave Front Sensing}

\author[a]{Olivier Fauvarque}
\author[a]{Benoit Neichel}
\author[a,b]{Thierry Fusco}
\author[a,b]{Jean-Francois Sauvage}
\author[a]{Orion Girault}

\affil[a]{Aix Marseille Univ, CNRS, LAM, Laboratoire d'Astrophysique de Marseille, Marseille, France}
\affil[b]{ONERA--the French Aerospace Laboratory, F-92322 Ch$\hat{\text{a}}$tillon, France}

\authorinfo{Further author information, send correspondence to \\Olivier Fauvarque : olivier.fauvarque@lam.fr\\Benoit Neichel : benoit.neichel@lam.fr\\}

\begin{document}
\maketitle

\begin{abstract}
We introduce in this article a general formalism for Fourier based wave front sensing. To do so, we consider the filtering mask as a free parameter. Such an approach allows to unify sensors like the Pyramid Wave Front Sensor (PWFS) and the Zernike Wave Front Sensor (ZWFS). In particular, we take the opportunity to generalize this two sensors in terms of sensors' class where optical quantities as, for instance, the apex angle for the PWFS or the depth of the Zernike mask for the ZWFS become free parameters.  In order to compare all the generated sensors of this two classes thanks to common performance criteria, we firstly define a general phase-linear quantity that we call meta-intensity. Analytical developments allow then to split the perfectly phase-linear behavior of a WFS from the non-linear contributions making robust and analytic definitions of the sensitivity and the linearity range possible. Moreover, we define a new quantity called the SD factor which characterizes the trade-off between these two antagonist quantities. These developments are generalized for modulation device and polychromatic light. A non-exhaustive study is finally led on the two classes allowing to retrieve the usual results and also make explicit the influence of the optical parameters introduced above.
\end{abstract}

\section{Introduction}\label{sec:intro}

By placing amplitude or phase masks in a focal plane, it is possible to filter the light from a pupil plane to another. Those masks are able, in particular, to transform incoming phase fluctuations into intensity variations on a detector.

Such optical designs are thus particularly relevant in the context of Wave Front Sensing, especially for the Adaptive Optics (AO). Moreover, Fourier based Wave Front Sensors (WFS) have many advantages compared to others Wave Front Sensors as for example the Shack-Hartmann in terms of, for instance, noise propagation or sampling flexibility.

The historical example of those Fourier based WFSs, dating from 1858, is the famous Foucault's knife. Ragazzoni generalized this physical concept with the Pyramid WFS \cite{Ragazzoni1996} which consists in using a 4-faces pyramidal mask in the focal plane. Variations of this concept regularly appear in the literature and define new WFSs: Akondi et al.  \cite{Akondi2014} and Vohnsen et al.\cite{Vohnsen2011} tested new pyramidal masks by changing the number of the faces of the pyramid; we recently proposed the Flattened Pyramid WFS \cite{Fauvarque2015} which introduces a new way to use the pyramid mask by reducing its apex angle. 

Using another physical concept based on phase contrast method, Zernike introduced the Zernike WFS \cite{zer1934} where the Fourier mask, completely transparent, has a circular depression in its midst allowing to create interference between the spatial frequencies of the incoming phase.

The essential purpose of this article is to merge all this Fourier based WFSs under a same mathematical formalism in order to build robust and relevant criteria allowing to compare their performance in the context of AO Wave Front Sensing. Such an approach has been described for instance in the context of phase masks in astronomy regarding interferometry and coronagraphy (Dohlen \cite{Dohlen2004}). 

In the first section of this article, we will present an original interpretation of the Fourier filtering technique thanks to the \textbf{focal plane tessellation formalism}. Such an approach will allow to describe all the masks involved in the AO Wave Front Sensing in an unique mathematical framework. 
We will secondly define an unified post-processing in order to create, from the detector output signal, a quantity called \textbf{meta-intensity} which will be linear with the phase. 
Thanks to an analytical development of the meta-intensity for any Fourier mask, we will, in the third part, 
define rigorously the \textbf{sensitivity} and the \textbf{linearity range} depending only on the choice of the Fourier mask used to do Wave Front Sensing. Moreover, we will show why these two performance criteria are inevitably antagonist thanks to a third criterion, the \textbf{SD factor}, which will quantify the trade-off between them. We will also give some hints regarding to the optimization of WFSs by considering masks as free-form objects. 
We will then show, in the fourth section, how it is possible to generalize these results in the context of a \textbf{modulation} in the pupil plane upstream to the filtering mask.
In the fifth and sixth sections, we will apply all these mathematical developments on the PWFS and the ZWFS. 
If in the four first sections, the light will be considered as monochromatic, a last section will be dedicated to the case of \textbf{polychromatic light}. In particular, we will show how our formalism allows to define the chromaticity of a sensor.

\section{Wave Front Sensing and Fourier Filtering}\label{opt_fou}

\subsection{Optical system}

We consider the optical system of figure \ref{ff_bench}.  The first plane contains the pupil and a focusing device corresponding to a perfect telescope. The associated spatial variables of this plane are $x_p$ and $y_p$; the $p$ index refers to "\textbf{P}upil plane". The second plane contains the mask and an imaging lens. The spatial variables are $f_x$ and $f_y$, this plane corresponds to the space of spatial \textbf{F}requencies
. The detector is placed in the third plane. This one is conjugated with the first pupil plane. The associated spatial variables  are $x_d$ and $y_d$ for "\textbf{D}etector".
\begin{figure}
\centering
\psscalebox{0.75 0.75} 
{\begin{Large}
\begin{pspicture}(0,-4.0783334)(16.7,4.0783334)
\definecolor{colour0}{rgb}{0.07450981,0.22745098,0.95686275}
\rput(8.756667,-0.69){point}
\rput(8.756667,-0.29){Focal}
\rput(3.99,3.9433333){Equivalent mirror focus +}
\rput(3.99,4.3433334){Incident EM field +}
\rput(3.99,3.5433334){\textcolor{blue}{Modulation}}
\rput(9.39,3.5433333){Imaging Lens}
\rput(9.39,3.9433334){\textcolor{red}{Focal mask \textbf{m}} +}
\psline[linecolor=black, linewidth=0.04](14.9,3.1433332)(14.9,-2.4566667)
\rput(14.1,-3.9233334){Detector's plane}
\psline[linecolor=black, linewidth=0.02, arrowsize=0.05291666666666667cm 2.0,arrowlength=1.4,arrowinset=0.0]{<->}(9.3,-3.2566667)(14.9,-3.2566667)
\rput(12.179999,-2.8566666){$f$}
\rput(6.579999,-2.8566666){$f$}
\rput(9.779999,2.9433334){$f/2$}
\rput(4.1799994,2.9433334){$f$}
\psline[linecolor=colour0, linewidth=0.04](1.7,-1.6566668)(3.7,-1.6566668)(14.9,2.3433332)
\psline[linecolor=colour0, linewidth=0.04](1.7,2.3433332)(3.670149,2.3433332)(14.9,-1.6566668)
\psline[linecolor=black, linewidth=0.04, arrowsize=0.05291666666666667cm 2.0,arrowlength=1.4,arrowinset=0.0]{<->}(9.3,3.1433332)(9.3,-2.4566667)
\psline[linecolor=black, linewidth=0.04, arrowsize=0.05291666666666667cm 2.0,arrowlength=1.4,arrowinset=0.0]{<->}(3.9,3.1433332)(3.9,-2.4566667)
\psline[linecolor=black, linewidth=0.04](3.7,-1.6566668)(3.7,-2.6566668)
\psline[linecolor=black, linewidth=0.04](3.7,3.3433332)(3.7,2.3433332)
\psline[linecolor=black, linewidth=0.02, arrowsize=0.05291666666666667cm 2.0,arrowlength=1.4,arrowinset=0.0]{<->}(3.7,-3.2566667)(9.3,-3.2566667)
\psline[linecolor=black, linewidth=0.04](3.3,-1.6566668)(4.1,-1.6566668)
\psline[linecolor=black, linewidth=0.04](3.3,2.3433332)(4.1,2.3433332)
\rput(14.79,3.9433333){Detector +}
\rput(14.79,3.5433333){\textcolor{blue}{Synchronization}}
\rput(9.256667,-3.9233334){Fourier plane}
\rput(3.8566668,-3.9233334){Pupil plane}
\rput(0.79,-2.79){light}
\rput(0.79,-2.39){collimated}
\rput(0.79,-1.9899999){Coherent}
\psline[linecolor=black, linewidth=0.04, arrowsize=0.05291666666666667cm 2.0,arrowlength=1.4,arrowinset=0.0]{->}(0.25666663,-1.4033333)(1.4566667,-1.4033333)
\psline[linecolor=black, linewidth=0.04, arrowsize=0.05291666666666667cm 2.0,arrowlength=1.4,arrowinset=0.0]{->}(0.25666663,-1.0033333)(1.4566667,-1.0033333)
\psline[linecolor=black, linewidth=0.04, arrowsize=0.05291666666666667cm 2.0,arrowlength=1.4,arrowinset=0.0]{->}(0.25666663,-0.6033333)(1.4566667,-0.6033333)
\psline[linecolor=black, linewidth=0.04, arrowsize=0.05291666666666667cm 2.0,arrowlength=1.4,arrowinset=0.0]{->}(0.25666663,-0.2033333)(1.4566667,-0.2033333)
\psline[linecolor=black, linewidth=0.04, arrowsize=0.05291666666666667cm 2.0,arrowlength=1.4,arrowinset=0.0]{->}(0.25666663,0.1966667)(1.4566667,0.1966667)
\psline[linecolor=black, linewidth=0.04, arrowsize=0.05291666666666667cm 2.0,arrowlength=1.4,arrowinset=0.0]{->}(0.25666663,0.59666675)(1.4566667,0.59666675)
\psline[linecolor=black, linewidth=0.04, arrowsize=0.05291666666666667cm 2.0,arrowlength=1.4,arrowinset=0.0]{->}(0.25666663,0.9966668)(1.4566667,0.9966668)
\psline[linecolor=black, linewidth=0.04, arrowsize=0.05291666666666667cm 2.0,arrowlength=1.4,arrowinset=0.0]{->}(0.25666663,1.3966668)(1.4566667,1.3966668)
\psline[linecolor=black, linewidth=0.04, arrowsize=0.05291666666666667cm 2.0,arrowlength=1.4,arrowinset=0.0]{->}(0.25666663,1.7966667)(1.4566667,1.7966667)
\psline[linecolor=black, linewidth=0.04, arrowsize=0.05291666666666667cm 2.0,arrowlength=1.4,arrowinset=0.0]{->}(0.25666663,2.1966667)(1.4566667,2.1966667)
\psline[linecolor=black, linewidth=0.02](15.3,0.3433334)(1.5,0.3433334)
\psbezier[linecolor=black, linewidth=0.04, arrowsize=0.05291666666666667cm 2.0,arrowlength=1.4,arrowinset=0.0]{->}(8.3,1.7433335)(8.3,1.2633334)(8.7,1.0233334)(9.1,0.5433334)
\rput[bl](8.1,1.9433334){\textbf{m}}
\end{pspicture}\end{Large}
}
\caption{Schematic view (in 1D) of a Fourier Filtering optical system. \label{ff_bench}}
\end{figure}
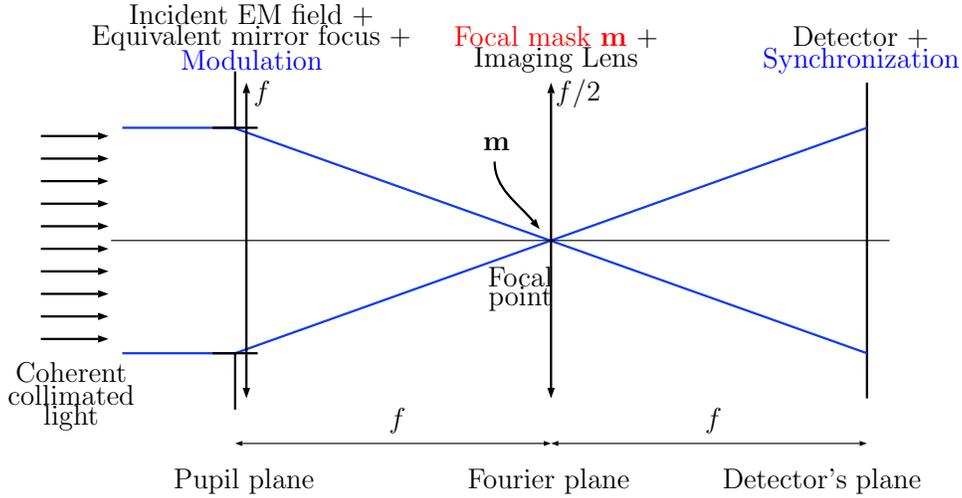\\
The incoming Electro Magnetic (EM) field can be written:
\begin{equation*}
\psi_p(x_p,y_p,\lambda) = \sqrt{n(\lambda)} \mathbb{I}_P(x_p,y_p) exp \left( \frac{2\imath\pi}{\lambda}\Delta(x_p,y_p) \right)
\end{equation*}
where $\lambda$ is the wavelength of the incoming light, $n(\lambda)$ is the number of photons by unit area at the wavelength $\lambda$, $\mathbb{I}_P$ is the indicator function of the pupil. $\Delta$ is the optical path difference created by atmospheric turbulence or any other sources of perturbation. \\
The mask has for transparency function $m(f_x,f_y,\lambda)$.  As a complex quantity, it can be decomposed into two terms:
\begin{equation*}
m(f_x, f_y,\lambda) = a(f_x,f_y,\lambda) ~~exp\left(\frac{2\imath\pi}{\lambda}OS(f_x,f_y,\lambda)\right) 
\end{equation*}
The function $a$ codes the amplitude filtering. Since the masks are passive, $a \in [0,1]$. The phase term is directly coded by the "Optical Shape" of the mask. This quantity "$OS$" depends on the optical indexes (and consequently on the wavelength for the refractive material) and on the geometric shape of the mask.  
Via the Fresnel optical formalism, it is possible to write the EM field in the detector plan :
\begin{equation}
\psi_d(x_d,y_d,\lambda) \propto \iint \frac{d x_p dy_p}{f^2} ~~\psi_p(x_p-x_d,y_p-y_d,\lambda)
\iint\frac{df_xdf_y}{\lambda^2}~~m(f_x,f_y,\lambda)~\exp \left(-\frac{2\imath\pi}{f\lambda}\left[x_p f_x+y_p f_y\right]\right) \label{grosseq}
\end{equation}

For the sake of clarity, we will consider in the first sections that the incoming light is monochromatic. The wavelength will thus be set at $\lambda_0$. Section \ref{pooly} will be dedicated to the polychromatic light case. 
The incoming EM field and the mask have the following expressions:
\begin{equation*}
\psi_p(x_d,y_d) = \sqrt{n}~\mathbb{I}_P(x_d,y_d) exp \left( \imath\phi(x_d,y_d) \right)
\end{equation*}
\begin{equation}
m(f_x, f_y) = a(f_x,f_y) ~~exp\left(\frac{2\imath\pi}{\lambda_0}  OS(f_x,f_y)\right) \label{maskkkkk}
\end{equation}
where $\phi = 2\pi \Delta / \lambda_0$ is the perturbed phase at the considered wavelength. The monochromatic assumption and the associated notations allows then to simplify equation \ref{grosseq} into the following form:
\begin{equation}
\psi_d = \psi_p \star \mathcal{F}[m] \label{eq_fonda}
\end{equation}
where $\mathcal{F}$ is the 2D Fourier Transform (2DFT) defined in appendix \ref{notations}. The interesting point about equation \ref{eq_fonda} is that the contributions of the pupil and the mask are clearly split: the EM field in the detector's plane $\psi_d$ is the convolution of the incoming field $\psi_p$ with the Fourier transform of the mask $\mathcal{F}[m]$. 

Moreover, since the Fourier transform is a bijective application, the quantity $\mathcal{F}[m]$ completely characterizes the mask. In other words, looking at $\mathcal{F}[m]$ or at $m$ itself is mathematically equivalent. Fraunhofer's diffraction also says that the $\mathcal{F}$ operator allows to go from a focal plane to a pupil plane. As a consequence, $\mathcal{F}[m]$ may be seen as the propagated EM field when the diffractive object is the mask $m$.  

One notes that equation \ref{eq_fonda} is not exact. Indeed, the optical design described above has a magnification equals to -1. This appears clearly when no mask is inserted in the focal plane. As a consequence, the term $\psi_d$ should be replaced by its symmetric, i.e. $\mathcal{S}[\psi_d]$ where $\mathcal{S}$ is the symmetric operator defined in appendix \ref{notations}. However, since this symmetric operation does not impact the wave front sensing, we will not consider it. 

The signal which is effectively obtained on the detector is the intensity associated to the EM field in the detector's plane $\psi_d$:
%
\begin{equation}
I = |\psi_p \star \mathcal{F}[m]|^2  \label{nl}
\end{equation}


\subsection{Considered masks in the context of Wave Front Sensing}

In this part, we will give a general interpretation of the effect of Fourier Filtering masks in the context of the Wave Front Sensing. The essential objective is to code the phase by using the incoming flux. Subsequently, the mask has to split and extract this information. These operations are done in the focal plane which is the Fourier Space associated to the Pupil plane. The area close to the optical axis contains the low spatial frequencies of the incoming EM field whereas the remote zones codes the high spatial frequencies. 

These physical results have been widely exploited to remove some spatial frequencies from images. We can for example mention the Abbe experiment and the Schlieren photography introduced by Toepler \cite{Toe64}. In our case, the use of opaque masks is not relevant since the reference sources are often faint. We therefore consider only masks having pure phase transparency functions. Mathematically, it means that the function $a$ in equation \ref{maskkkkk} equals to 1 in the entire focal plane.
Having said that, we will now see how it is possible to split and then extract the spatial frequencies information thanks to a Fourier mask. 

The first step is to choose a tessellation $(\Omega_i)$ of the Fourier plane. Mathematically, it means that:
\begin{equation*}
\cup_i \Omega_i = \mathbb{R}^2 \text{~~~and~~~} \Omega_i \cap \Omega_j = \emptyset \text{~~~if~~~}i \neq j
\end{equation*}
The index $i$ goes in a finite or a countable set. Each elements of this tessellation allows to select a certain part of the spatial frequencies. The second step consists in separating the spatial frequencies contained in each elements $\Omega_i$ thanks to a tip and a tilt generated via a slope (or an angle) in the optical shape of the mask.
If several $\Omega_i$ have the same rejection angle, it is possible to do interference between them thanks to local pistons. This corresponds, in terms of optical shape, to local thickness differences.(Obviously, it is possible to add the next shapes (focus, astigmatism, etc.) by playing with the optical shape of each $\Omega_i$ but we choose to not explore this level of complexity in this article.) Mathematically, every considered masks can thus be written: 
\begin{equation}
m(f_x,f_y)=\sum_i \mathbb{I}_{\Omega_i}(f_x,f_y) \exp \left(\frac{2\imath\pi}{\lambda_0} (\delta_i+f_x\alpha_i+f_y\beta_i)\right)\label{masss}
\end{equation}
where $\delta_i$ codes the local piston and $(\alpha_i,\beta_i)$ the local tip/tilt corresponding to the rejection angles. $ \mathbb{I}_{\Omega_i}$ is the indicator function of $\Omega_i$. 
To summarize, it is possible to describe any kind of mask used in the context of wave front sensing via the following  set of parameters:
\begin{equation}
\{ \Omega_i, \delta_i,\alpha_i,\beta_i \} _i \label{tess_param}
\end{equation}
This set will be called the \emph{tessellation parameters} of the considered mask. With this formalism, it is now possible  to write the 2DFT corresponding to the expression of the general transparency function (equation \ref{masss}):
\begin{equation}
\mathcal{F}[m](x_d,y_d)=\sum_i \exp \left(\frac{2\imath\pi\delta_i}{\lambda_0}\right) \mathcal{F}[\mathbb{I}_{\Omega_i}](x_d-f\alpha_i,y_d-f\beta_i) \label{CIR}
\end{equation}
where $f$ is the focal of the imaging lens. 

\subsection{2D Fourier Transform of the classical tessellations}

\begin{figure}[h]
\begin{minipage}[c]{0.45\linewidth}
\begin{center}
\psscalebox{1.0 1.0} 
{
\begin{pspicture}(0,-1.8)(3.1069584,1.8)
\psline[linecolor=black, linewidth=0.04, arrowsize=0.05291666666666667cm 2.0,arrowlength=1.4,arrowinset=0.0]{<-}(1.6,1.4)(1.6,-1.8)
\psline[linecolor=black, linewidth=0.04, arrowsize=0.05291666666666667cm 2.0,arrowlength=1.4,arrowinset=0.0]{->}(0.0,-0.2)(3.2,-0.2)
\rput[bl](0.5,0.6){$\Omega^{-+}$}
\rput[bl](0.5,-1.0){$\Omega^{--}$}
\rput[bl](2.1333334,-1.0){$\Omega^{+-}$}
\rput[bl](2.1333334,0.6){$\Omega^{++}$}
\rput[bl](1.4,1.6){$f_y$}
\rput[bl](3.4,-0.3){$f_x$}
\end{pspicture}
}
\end{center}
\end{minipage}~~
\begin{minipage}[c]{0.45\linewidth}
\begin{center}
\psscalebox{1.0 1.0} 
{
\begin{pspicture}(0,-1.8)(3.1069584,1.8)
\definecolor{colour2}{rgb}{0.9607843,0.9607843,0.9607843}
\psline[linecolor=black, linewidth=0.02, arrowsize=0.05291666666666667cm 2.0,arrowlength=1.4,arrowinset=0.0]{<-}(1.6,1.4)(1.6,-1.8)
\psline[linecolor=black, linewidth=0.02, arrowsize=0.05291666666666667cm 2.0,arrowlength=1.4,arrowinset=0.0]{->}(0.0,-0.2)(3.2,-0.2)
\pscircle[linecolor=black, linewidth=0.04, fillstyle=solid,fillcolor=colour2, dimen=outer](1.6,-0.2){0.8}
\rput[bl](1.4,0.0){$\Omega_\rho$}
\psline[linecolor=black, linewidth=0.02, arrowsize=0.05291666666666667cm 2.0,arrowlength=1.4,arrowinset=0.0]{<->}(1.6,-0.2)(2.2,-0.7)
\rput[bl](1.65,-0.7){$\rho$}
\rput[bl](1.4,1.6){$f_y$}
\rput[bl](3.4,-0.3){$f_x$}
\rput[bl](2.4,0.8){$\bar{\Omega}_\rho$}
\end{pspicture}
}
\end{center}
\end{minipage}
\vspace*{0.2cm}
\caption{Two classical tessellations. Cartesian splitting (left insert) and polar splitting (right insert).\label{cart}}
\end{figure}
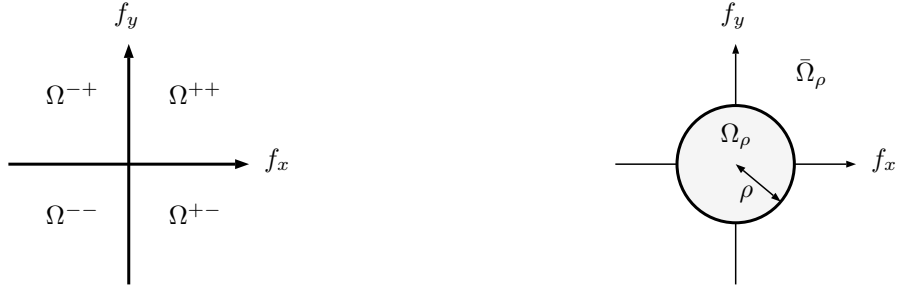

Although all the tessellations of the Fourier plan are \emph{a priori} acceptable, two of them play a significant role for the existing WFSs.\\
The first one consists in splitting the plane into 4 elements following the Cartesian coordinate system (left insert of figure \ref{cart}).
The second one is a polar splitting of the Fourier plane. This tessellation has a parameter $\rho$ which codes the size of the central circle (right insert of figure \ref{cart}).

\noindent From these definitions, one can write the indicator functions of each part of the tessellation and then get their 2DFT. We detail this indicator function for the element~$\Omega^{++}$:
\begin{equation*}
\mathbb{I}_{\Omega^{++}}(f_x,f_y) = \Theta(f_x)\Theta(f_y) 
\end{equation*}
where $\Theta$ is the Heaviside function. Hence its 2DFT equals to:
\begin{equation}
\mathcal{F}[\mathbb{I}_{\Omega^{++}}](x_d,y_d) = \frac{1}{4}\left(\delta(x_d)\delta(y_d)-\frac{1}{\pi^2 x_d y_d}\right)-\frac{\imath}{4}\left(\frac{\delta(x_d)}{\pi y_d} +\frac{\delta(y_d)}{\pi x_d}  \right)  \label{26}
\end{equation}
For the other parts of the Cartesian tessellation, we get:
\begin{equation}
\mathcal{F}[\mathbb{I}_{\Omega^{-+}}](x_d,y_d) = \frac{1}{4}\left(\delta(x_d)\delta(y_d)+\frac{1}{\pi^2 x_d y_d}\right)-\frac{\imath}{4}\left(\frac{\delta(x_d)}{\pi y_d} -\frac{\delta(y_d)}{\pi x_d}  \right) \label{27} 
\end{equation}
\begin{equation}
\mathcal{F}[\mathbb{I}_{\Omega^{--}}](x_d,y_d) = \frac{1}{4}\left(\delta(x_d)\delta(y_d)-\frac{1}{\pi^2 x_d y_d}\right)+\frac{\imath}{4}\left(\frac{\delta(x_d)}{\pi y_d} +\frac{\delta(y_d)}{\pi x_d}  \right) \label{28} 
\end{equation}
\begin{equation}
\mathcal{F}[\mathbb{I}_{\Omega^{+-}}](x_d,y_d) = \frac{1}{4}\left(\delta(x_d)\delta(y_d)+\frac{1}{\pi^2 x_d y_d}\right)+\frac{\imath}{4}\left(\frac{\delta(x_d)}{\pi y_d} -\frac{\delta(y_d)}{\pi x_d}  \right) \label{29} 
\end{equation}
For the polar tessellation, the indicator function  $\mathbb{I}_{\Omega_\rho}(f_r,f_\theta)$ equals to $\Theta(\rho-f_r)$. Its 2DFT is:
\begin{eqnarray}
\mathcal{F}[\mathbb{I}_{\Omega_\rho}](r_d, \theta_d)= \frac{\rho}{r_d} J_1(2\pi \rho r_d) \label{polar_tess}
\end{eqnarray}
where $J_{\alpha}$ are the first kind Bessel functions.

\subsection{Other WFSensors}
We mention here that other tessellations may be relevant in the WFSensing context, especially regarding to the case of the Point Diffraction Interferometer introduced by Smartt \cite{smartt1974}. Moreover, this formalism easily extends to the Optical differentiation WaveFront Sensor introduced by Oti \cite{Oti03} and also describes the first stage of coronagraphic systems.

\section{Meta-intensity definitions}

In this section, we introduce the basic elements needed to construct, from the intensity on the detector, a numerical quantity, called \emph{meta-intensity} and written $mI$, which consists in the linear response to an incoming turbulent phase. 

\subsection{Reference phase}


The phase seen by the WFS is the sum of the turbulent phase induced by the atmosphere and the static aberrations of the wave front sensing path. 
Mathematically, we can split the incoming phase into two terms:
\begin{equation}
\phi = \phi_r + \phi_t\label{phii}
\end{equation}
where $\phi_t$ is the turbulent phase and $\phi_r$ the static reference phase. $\phi_r$ may also be seen as the operating point of the WFS.


\subsection{Towards the linearity}

The goal of the meta-intensity is to transform the intensity on the detector into a quantity which will be linear with the turbulent phase around the reference phase. Moreover, the $mI$ has also to be independent to the flux $n$. Mathematically, such conditions are written:
\begin{equation*}
mI(\phi_t+a\Phi_t) = mI(\phi_t)+a~mI(\Phi_t)~~~~~~ \forall \phi_t,\Phi_t \in \text{Phase space and~~} \forall a\in\mathbb{R}
\end{equation*}
We make use of the power series of exponential function and the Cauchy product laws to develop the squared module of equation \ref{nl} and get an expression of the intensity depending on the successive powers of the turbulent phase: 
\begin{equation}
\hspace{-0.25cm}I(\phi,n) = I(\phi_r+\phi_t,n) = n\sum_{q=0}^{\infty}\frac{(-\imath)^q}{q!} \sum_{k=0}^{q} (-1)^{k} {q \choose k} \phi^{k\star}_t \overline{\phi^{q-k\star}_t}~~~~~~\text{where}~~~ \phi^{k\star}_t \hat{=} (\mathbb{I}_p ~e^{\imath \phi_r}~\phi_t^k) \star \mathcal{F}[m] \label{CauchyLaww}
\end{equation}
The complex quantity $\phi^{k\star}_t$ may be considered as a $k^{th}$-moment of the turbulent phase through the mask around the static reference phase $\phi_r$. 

By looking at equation \ref{CauchyLaww}, we note that a way to make $I$ independent on the flux is to divide it by the spatial average incoming flux $n$. This operation is easy in practice since $n$ is proportional to the total flux on the detector. 
The first step of the post-processing to build the meta-intensity is thus to \textbf{normalize} the intensity by the spatial average flux $n$.

Regarding the linearity with respects to the phase, one can explicitly develop equation \ref{CauchyLaww}. This is done below for $q=0$ to 2: 
\begin{center}
Constant term, $q=0$:
\end{center}
\begin{equation}
I_c~~~\hat{=}~|\mathbb{I}_P ~e^{\imath \phi_r} \star \mathcal{F}[m]|^2\label{mIc}
\end{equation}
\begin{center}
Linear term, $q=1$:
\end{center}
\begin{equation}
I_l(\phi_t)~~~\hat{=}~2\Im[(\mathbb{I}_P~e^{\imath \phi_r} \star \mathcal{F}[m])(\overline{\mathbb{I}_P~e^{\imath \phi_r}\phi_t \star \mathcal{F}[m]})]\label{mIl}
\end{equation}
\begin{center}
Quadratic term, $q=2$:
\end{center}
\begin{equation}
I_q(\phi_t)~~~\hat{=}~|\mathbb{I}_P~e^{\imath \phi_r} \phi_t \star \mathcal{F}[m]|^2-\\
\Re[(\mathbb{I}_P~e^{\imath \phi_r} \star \mathcal{F}[m])(\overline{\mathbb{I}_P~e^{\imath \phi_r} \phi^2_t \star \mathcal{F}[m]})]\label{mIq}
\end{equation}
The only term linear with the turbulent phase $\phi_t$ is the $q=1$ term that we call \emph{linear intensity} $I_l$. Ideally, and in order to maximize the linearity, one would want to minimize the other terms, i.e. $q=0$ and $q \geq 2$.
The constant intensity $I_c$ ($q=0$) corresponds to the normalized intensity on the detector when the phase equals to the reference phase, i.e. $I(\phi_r,n)/n$. Removing such a term can thus be done thanks to a calibration path.
With regard to the $q \geq 2$ terms, they are unfortunately impossible to remove but we still consider that the second step to build $mI$ consists in this \textbf{return-to-reference}. Mathematically, the meta-intensity $mI$ is thus defined as:
\begin{equation}
mI(\phi_t)= \frac{I(\phi_r+\phi_t,n)-I(\phi_r,n)}{n}\label{MI}
\end{equation}
This definition corresponds in fact in the easiest way to define a linear quantity for any type of sensor. 
Finally, it might be relevant to precise that it exits a lot of other processes which can be applied to the meta-intensities after this two fundamental first steps (normalization, return-to-reference) but it is a vast topic that will be tackled in a forthcoming article (Fauvarque et al. \cite{Fauvarque2016prep}). 

%
%

\subsection{Linear and effective meta-intensities}

The previous post-processing allows to build, \textbf{in practice}, the meta-intensity from the intensity on the detector and a reference intensity. Unfortunately, we saw that equation \ref{MI} is not perfectly linear with the turbulent phase since $mI$ still contains the quadratic and next high-order moments of the phase. In the following, we will distinguish the effective behavior of a WFS from its ideal one. Equation \ref{MI} will define the \emph{effective} meta-intensity while the restriction of the phase power series development to its first term, i.e. to the linear intensity $I_l$, will define the \emph{linear meta-intensity} (see equation \ref{mIl}). 
We note that this linear meta-intensity cannot be strictly deduced from the intensity on the detector but easily computed by numerical simulations. It corresponds to the behavior which would have a perfectly linear WFS.  

As side comment, another interpretation of the linear and quadratic meta-intensities can be deduced from the derivatives of the normalized intensity with respect to the amplitude phase around the reference phase, i.e.:
\begin{eqnarray*}
I_l(\phi_t)&=  \frac{1}{n} \left. \frac{d I(\phi_r+a\phi_t,n)}{d a} \right|_{a=0}\\ 
I_q(\phi_t)&=  \frac{1}{2 n} \left. \frac{d^2 I(\phi_r+a\phi_t,n)}{d a^2} \right|_{a=0}
\end{eqnarray*}
This result is not surprising regarding the calculation of  equation \ref{CauchyLaww} which is the Taylor's development of the intensity around $\phi_r$. 

\section{Sensitivity and Linearity Range}
In this section, we define the sensitivity and the linearity range of a WFS regarding to an incoming turbulent phase. From now, this turbulent phase will be normalized regarding the 2-norm (i.e. the RMS norm). $\phi_t$ in the light of the previous developments. 

Due to the fact that the \textbf{sensitivity} only makes sense in the linear regime of the WFS, we will use the \emph{linear meta-intensity} $I_l$ to define it. On the other hand, the \textbf{linearity range} will be calculated thanks to the study of the variation of the distance between the effective and the linear meta-intensities when the input phase amplitude is varying. 

\subsection{Sensitivity}\label{sensitivity}

The perfectly linear response to the normalized incoming phase $\phi_t$ is obtained thanks to the expression of the \emph{linear} intensity (equation \ref{mIl}) $I_l(\phi_t)$.

We choose to define as "the sensitivity regarding to $\phi_t$" the 2-norm of this linear response:
\begin{equation}
s(\phi_t) = ||I_l(\phi_t)||_2 \label{sensii}
\end{equation}
Such a definition is in fact consistent with the method introduced by Rigaut et al.\cite{Rigaut92} to estimate the noise propagation associated to a WFS. The Rigaut and Gendron method consists in looking at the diagonal terms of the matrix $(^t\mathcal{M}\mathcal{C}_B^{-1}\mathcal{M})^{-1}$, where $\mathcal{M}$ is the interaction matrix, i.e. the matrix which contains the output meta-intensities for an input phase basis, and $\mathcal{C}_B$ is the noise covariance matrix.
Physically, these values correspond to the noise propagation coefficients for a detector or photon noise regime depending the nature of $\mathcal{C}_B$. 

In the case of a diagonal and constant noise covariance matrix, i.e. $\mathcal{C}_B=\sigma^2_\text{noise}\times Id$, the noise propagation coefficients for the turbulent phase $\phi_t$, called $\sigma_\text{WFS}^2(\phi_t)$ in Rigaut et al. are related to the sensitivity definition introduced above $s(\phi_t)$ via the following equation: 
\begin{equation*}
\sigma_\text{WFS}^2(\phi_t) = \sigma_\text{noise}^2~s(\phi_t)^{-2}
\end{equation*}

\subsection{Linearity range}\label{LinRange}

\begin{figure}[htb]
\centering
\rotatebox{90}{\hspace{1.1cm}$G_{el}$~~~ Gap Effective/Linear }
\includegraphics[trim = 1cm 1.5cm 1.7cm 2cm, clip,width=8cm]{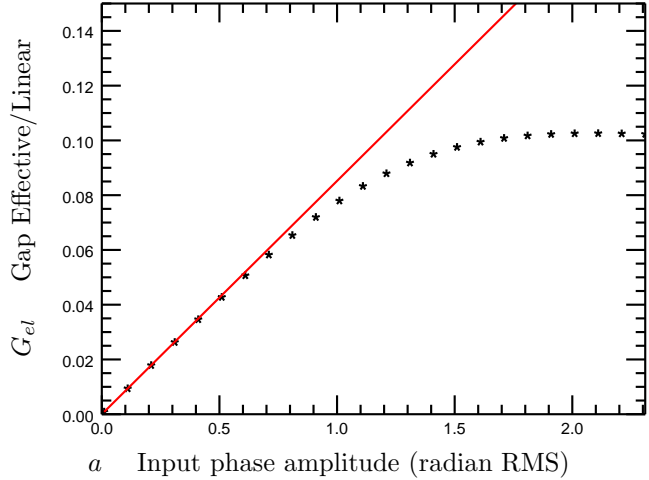}\\
$a$~~~ Input phase amplitude (radian RMS)
\caption{Typical graph ($*$) of the gap between the effective meta-intensity and the linear intensity inte depending on $a$, the input phase amplitude (equation \ref{LR}). The slope of the red line equals to the 2-norm of the quadratic intensity. The input phase is the vertical coma and the considered WFS is the Zernike WFS.  \label{GAP_le}}
\end{figure}

The linearity range quantifies the gap between the perfectly linear behavior of a WFS and its effective behavior.  Indeed, when the incoming phase amplitude increases the meta-intensity defined in equation \ref{MI} differs more and more from the linear intensity (equation \ref{mIl}). We call the mathematical function which quantifies this deviation from linearity $G_{el}$ for "\textbf{G}ap between \textbf{e}ffective and \textbf{l}inear behaviors" and we define it by:
\begin{equation}
G_{el}(\phi_t,a) = \left|\left|\frac{mI(a\phi_t)}{a}- I_l(\phi_t)\right|\right|_2 \label{LR}
\end{equation}
where the variable $a$ codes the amplitude of the incoming phase. We note that $\phi_t$ is, once again, normalized regarding to the RMS norm. The black curve of figure \ref{GAP_le} shows the typical evolution of $G_{el}$ when $a$ increases. Two distinct regimes are observed: a linear growth for the lowest amplitude and then, a saturation regime. In the saturation regime, the intensity on the detector does not change anymore as the phase amplitude continues to grow.
The linear regime can be explained by looking at the analytic expression of $G_{el}$: 
\begin{eqnarray*}
\frac{mI(a\phi_t)}{a}- I_l(\phi_t) = &\sum\limits_{q=2}^{\infty} a^{q-1}\frac{(-\imath)^q}{q!} \sum\limits_{k=0}^{q} (-1)^{k}{q \choose k} \phi^{k\star}_t \overline{\phi^{q-k\star}_t}\\ &=a I_q(\phi_t) + a^2(...)
\end{eqnarray*}
The linear growth is thus directly linked to the quadratic intensity, more precisely the associated slope equals the 2-norm of $I_q$ (see the red curve of figure \ref{GAP_le}):
\begin{equation*}
G_{el}(\phi_t,a) = a||I_q(\phi_t)||_2 + a^2(...)
\end{equation*}
In others words, a small slope means that the effective meta-intensities slowly differs from the linear behavior for an increasing incoming phase amplitude ; the associated linear range is thus large.
We choose thus to define the linearity range regarding to the normalized phase mode $\phi_t$  as the inverse of the 2-norm of the quadratic term:
\begin{equation}
d(\phi_t) = \Big(||I_q(\phi_t)||_2\Big)^{-1} \label{dynaa}
\end{equation}
where the letter $d$ refers to the word "\textbf{d}ynamic" which is a synonym of linearity range.

Such a definition allows to calculate the linear range in an analytic way thanks to the expression of the quadratic term (equation \ref{mIq}).

\subsection{The SD Factor}

The previous developments allow us to give an explication about the fact that the sensitivity and the linearity range are antagonist quantities. By looking at their product, that we call the SD factor:   
\begin{equation}
s(\phi_t)d(\phi_t) = \Big(||I_l(\phi_t)||_2\Big)\Big(||I_q(\phi_t)||_2\Big)^{-1}
\end{equation}
we can see that such a quantity corresponds to the ratio between the norms of the first and the second derivatives of the intensity regarding to the turbulent phase. One can easily understand that it is difficult to increase the numerator while decreasing the denominator at the same time. 
The \emph{SD factor} is thus a relevant indicator of the \emph{trade-off between the sensitivity and the linearity range}.

The expressions of the linear and quadratic intensities (equations \ref{mIl} and \ref{mIq}) and the 2-norm definition allow to numerically optimize a WFS, i.e. its mask, by maximizing its SD factor. For a given turbulent phase $\phi_t$, we can thus consider a mask depending on a scalar optical parameter that we call $p$. The only constraint on this parameter is to define a mask which is transparent, i.e. $|m(f_x,f_y)|=1$ (or at least a passive one: $|m(f_x,f_y)|\leq 1$). As examples, the optical parameters of the tessellation formalism $(\delta, \alpha, \beta)$ are appropriate since they are variables for the Geometrical Shape of the mask.
The next step just consists in finding numerically the maximum of the function $s.d(\phi_t,m(p))$ regarding to the parameter $p$. 


Finally, one can modify the SD factor to give more or less importance to the sensitivity or the linearity range by introducing a power exponent $\eta$ in it:
\begin{equation*}
s^{\eta}d^{1/\eta} = \Big(||I_l(\phi_t)||_2\Big)^{\eta}.\Big(||I_q(\phi_t)||_2\Big)^{-1/\eta}
\end{equation*}
with $\eta \in \mathbb{R}_+^*$. For instance, $\eta=0.5$ confers more importance to the linearity range than to the sensitivity whereas $\eta=2$ represents the opposite case. The power exponent $\eta$ allows to consider different requirement specifications. 

\section{Modulation}\label{ttt}

The Fourier filtering may be coupled with an additional optical stage placed in the entrance pupil plane. Such a device creates an oscillating aberration and changes the shape and the size of the focal spot on the Fourier mask. The detector is synchronized with this regular oscillation in order to have one image for each aberration cycle. This system called \emph{modulation} allows to adjust the WFS's performance.

Historically, such a device has been introduced by Ragazzoni \cite{Ragazzoni1996} in order to improve the linearity range of the classical PWFS. Even if, to our knowledge, modulation is only used for the PWFS, we give in this section a definition for any kind of WFSs and show how to define a generalized sensitivity and linearity range with such a device working.

\subsection{General definition of modulation}

The oscillating aberration introduced by the modulation device defines a closed path (or loop) in the phase space. Mathematically, it means that modulation may be defined in the following way: 
\begin{eqnarray*}
&\phi_m(s) = \displaystyle \sum_{k=0}^\infty m_k(s) \Phi_k &\text{~~~~with~~~~} \forall k~~ m_k(0) = m_k(1)  \\
&w(s) \text{~~~with~~~}  w(s)\geq 0, & w(0)=w(1), \int_0^1 w(s) ds = 1 
\end{eqnarray*}
where $s$ is the temporal variable normalized with respects to the duration of the modulation cycle $\tau$, i.e. $s=t/\tau$. The phase polynomials $\Phi_k$ describe a phase basis typically the Zernike basis. The functions $m_k(s)$ indicate the amplitudes of the phase modes used during modulation. $w(s)$ is the weighting function, it codes the time spent for each modulation phase $\phi_m(s)$. 

\subsection{Intensity on the detector}\label{refpha}

Thanks to these definitions, it is possible to write the modulated intensity, called $I_m$, on the detector. In particular, this one is the integral during a cycle of the intensity with an additional phase corresponding to the local modulation phase:
\begin{equation*}
I_m(\phi,n) =\int_0^1 I(\phi+\phi_m(s),n.w(s))ds 
\end{equation*} 
Such an equation is linked to the fact that modulation handles the light as an incoherent quantity. Moreover, remembering equation \ref{phii} it appears that 
the reference phase may be seen as a static modulation. In the following, we will assume that the modulation phase contains this reference phase. The generalized phase power series of the intensity becomes thus:
\begin{equation*}
\hspace{-0.25cm}\frac{I_m(\phi_t,n)}{n} = \sum_{q=0}^{\infty} \frac{(-\imath)^q}{q!} \sum_{k=0}^{q} (-1)^{k} {q \choose k}\int_0^1 w(s)ds\phi^{k\star}_t(s) \overline{\phi^{q-k\star}_t(s)}~~~~~\text{where}~~~ \phi^{k\star}_t(s) = (\mathbb{I}_p ~e^{\imath \phi_m(s)}~\phi_t^k) \star \mathcal{F}[m] 
\end{equation*}

\subsection{Modulated meta-intensity}

The first terms of the previous equation corresponds once again to the modulated constant, linear and quadratic terms:
\begin{equation*}
I_{mc} = \int_0^1 |\mathbb{I}_P ~e^{\imath \phi_m(s)} \star \mathcal{F}[m]|^2 w(s)ds 
\end{equation*}
\begin{equation*}
I_{ml}(\phi_t) = \int_0^1 2\Im[(\mathbb{I}_P~e^{\imath \phi_m(s)} \star \mathcal{F}[m])
(\overline{\mathbb{I}_P~e^{\imath \phi_m(s)}\phi_t \star \mathcal{F}[m]})]w(s)ds
\end{equation*}
\begin{equation*}
I_{mq}(\phi_t) = \int_0^1\left[|\mathbb{I}_P~e^{\imath \phi_m(s)} \phi_t \star \mathcal{F}[m]|^2\right]w(s)ds-\int_0^1 \left[\Re[(\mathbb{I}_P~e^{\imath \phi_m(s)} \star \mathcal{F}[m])(\overline{\mathbb{I}_P~e^{\imath \phi_m(s)} \phi^2_t \star \mathcal{F}[m]})]\right]w(s)ds
\end{equation*}
The method used previously  to construct a phase-linear quantity (at least in the small phases approximation) is thus still valid. The effective modulated meta-intensity, called $mI_m$, will be subsequently defined in the same way as equation \ref{MI}, i.e.: 
\begin{equation}
mI_m(\phi_t)= \frac{I_m(\phi_t,n)-I_m(0,n)}{n}\label{MIm}
\end{equation}
The definitions of the sensitivity and the dynamic stay as they were in equations \ref{sensii} and \ref{dynaa} under the condition to use now \textbf{modulated} linear and quadratic intensities $I_{ml}$ and $I_{mq}$.

\section{Application to the Pyramid WFS}

In this section, we apply the theoretical formalism developed above for the Pyramid WFSs. Without loss of generality, we assume in the next developments that the operating phase equals to the null phase. Ragazzoni \cite{Ragazzoni1996} introduced this sensor which is a generalization of the Foucault's Knife. The corresponding Fourier mask placed in the focal plane is a transparent squared pyramid. 
Its apex angle is called $\theta$. In its historical configuration, this mask creates in the detector's plane 4 pupil images containing each only one part of the spatial frequencies.

\subsection{Tessellation formalism}
\begin{figure}[htbp]
\centering

\includegraphics[width=4cm]{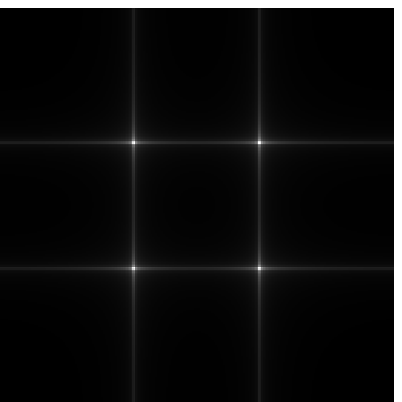}~~~~
\includegraphics[width=4cm]{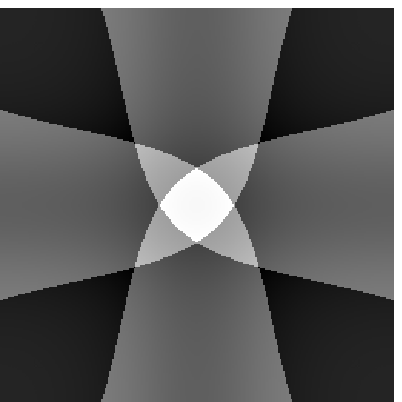}
\caption{Modulus (left) and argument (right) of the 2DFT with an apex angle which equals to $\frac{1.5D}{2f}$ where $D$ is the entrance pupil diameter. \label{CIR_pyr}}
\end{figure}

\begin{figure}[htbp]
\centering
\includegraphics[width=4cm]{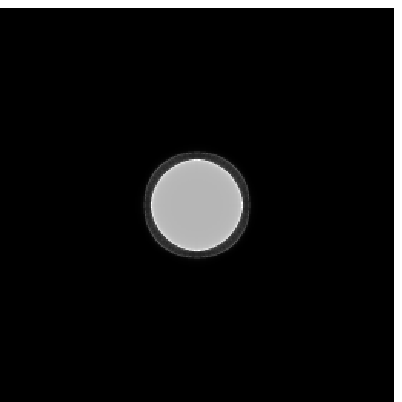}~~
\includegraphics[width=4cm]{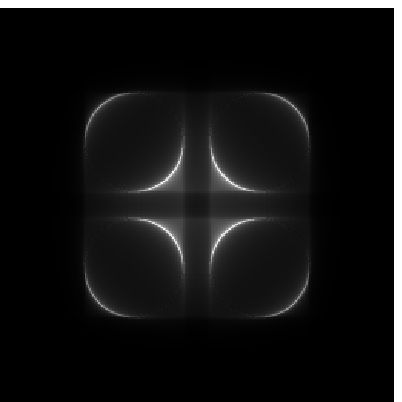}

\vspace{0.15cm}

\includegraphics[width=4cm]{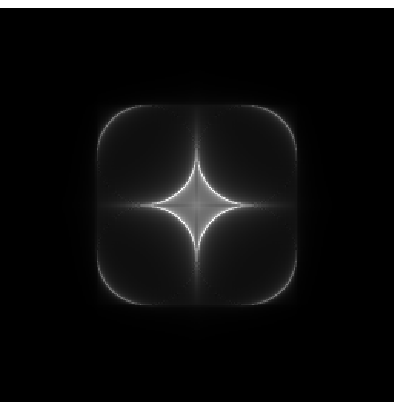}~~
\includegraphics[width=4cm]{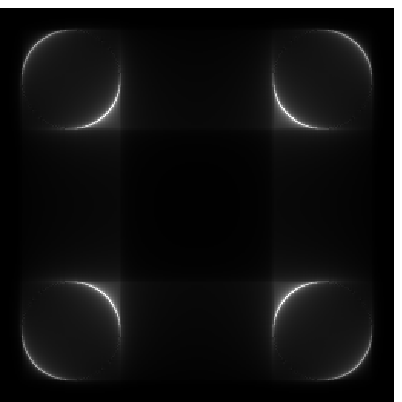}
\caption{Intensity on the detector for a circular pupil and a flat incoming phase. $\theta$ equals to  0.1, 1, 1.5 and 3$\frac{D}{2f}$.\label{Fc_ecart}}
\end{figure}

The pyramidal mask allows to illustrate the spatial frequencies separation induced by local tip/tilt OPDs. 
%
The tessellation parameters (i.e. $\Omega_i: \delta_i,\alpha_i,\beta_i$), associated to the transparency function of this mask that we note $m_\triangle$ are:
\begin{eqnarray*}
&\Omega^{-+} :~~0, -\theta, \theta ~~~~~ &\Omega^{++} :~~0, \theta, \theta \\
&\Omega^{--} :~~0, -\theta, -\theta ~~~~~~ &\Omega^{+-} :~~0,\theta, -\theta 
\end{eqnarray*}
The associated 2DFT is:
\vspace*{-0.2cm}
\begin{multline}
\mathcal{F}[m_\triangle](x_d,y_d)= 
\mathcal{F}[\mathbb{I}_{\Omega^{++}}](x_d-f\theta,y_d-f\theta)+\mathcal{F}[\mathbb{I}_{\Omega^{--}}](x_d+f\theta,y_d+f\theta)+\\
\mathcal{F}[\mathbb{I}_{\Omega^{-+}}](x_d+f\theta,y_d-f\theta)+\mathcal{F}[\mathbb{I}_{\Omega^{+-}}](x_d-f\theta,y_d+f\theta)
\label{pyrtt}
\end{multline}
Considering the 2DFT of each quadrant $\Omega_i$ (see equations \ref{26}, \ref{27}, \ref{28} and \ref{29}) we have with equation \ref{pyrtt}, an analytic formulation of the 2DFT of the Pyramid WFS mask. 
We can see, on figure \ref{CIR_pyr} the module and the argument of this quantity (equation \ref{pyrtt}).
Such quantities allows us to determine the imaging on the detector (top right insert of figure \ref{Fc_ecart}) since this intensity corresponds to the convolution between the entrance pupil and the 2DFT of the mask.
\noindent It may be interesting to study the different cases of this optical design with respects to the apex angle parameter: 

- The first one is when $\theta$ tends to infinity. It corresponds to the case of a reflective pyramid which creates 4 pupil images on 4 different detectors. This design is introduced by Wang et al.\cite{Wang10}. It allows to completely separate the 4 pupil images. Concretely, it means that the four 2DFTs are considered as independent: there is no cross talk between the spatial frequencies of the 4 quadrants $\Omega_i$. This hypothesis of no interferences between the pupil is usually done in theoretical approaches and in usual simulation algorithms.

- The second case is the more common one. It corresponds to an achromatic refractive pyramid. The 4 pupil images are created on a unique detector and $D/2 \leq f\theta$
where $D$ is the diameter of the entrance pupil and $f$ the focal of the imaging lens. It means that there is no overlap between these 4 images even if each pupil image contains a part of the spatial frequencies of the others quadrants due to the fact that the 2DFTs of the four $\Omega_i$ have not a compact support. This interference between the 4 pupil images is particularly marked in the midst area (see top right insert and bottom inserts of figure \ref{Fc_ecart}).

- We introduced the last case in the letter \cite{Fauvarque2015} with the Flattened Pyramid WFS. The idea of this sensor is to overlap the 4 pupil images by using a small angle: $0<\theta f<D/2$. (The case $\theta=0$ is obviously useless since it corresponds to the trivial mask.) This optical configuration is shown on top left insert of figure \ref{Fc_ecart}.

Before studying into more details these WFSs, one notes that it would be relevant to now consider the Pyramid not as a unique and "static" optical system but more like a \emph{class} including an infinite number of different pyramids. Indeed the angle of the apex $\theta$ is one optical parameter and it is possible to envisage another ones as, for example, the number of the faces or the modulation parameters.    

\subsection{Sensitivity, linearity range and SD factor}

\subsubsection{Apex angle}
\begin{figure}[htb]
\centering
\rotatebox{90}{\hspace{2.5cm}\textcolor{red}{\textbf{s}}~~~~ \textcolor{blue}{\textbf{s.d}}~~~~\textbf{d} }
\includegraphics[trim = 1cm 1.5cm 1.7cm 1.8cm, clip,width=8cm]{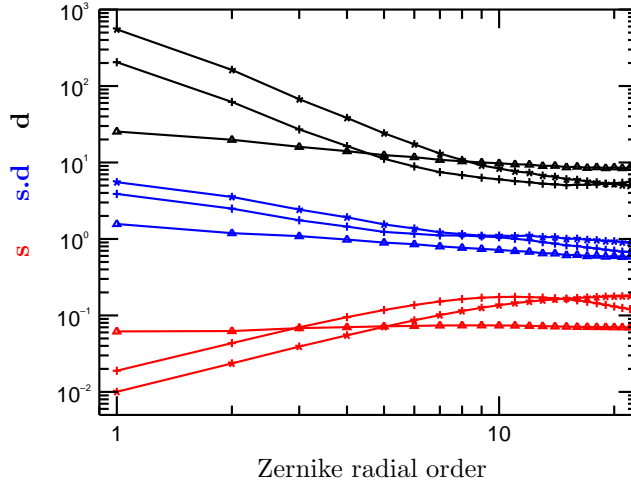}\\
\hspace{1.25cm}Zernike radial order
\caption{\textcolor{red}{\textbf{Sensitivity}}, \textbf{Linearity range} and \textcolor{blue}{\textbf{SD factor}} of the Pyramid WFS with respects to the spatial frequencies. \textbf{Apex angle} equals to 0.05 ($*$), 0.1 ($+$) and $3$ ($\triangle$) $\frac{2\theta f}{D}$. The phase basis corresponds to the 24 first Zernike radial orders.\label{SD_pyr}}
\end{figure}

Figure \ref{SD_pyr} shows the sensitivity, the linearity range and the SD factor (red, black and blue curves) with respect to the first 24 Zernike radial orders -- i.e. the spatial frequencies  -- for different apex angles. 
Physically, this optical parameter $\theta$ sets the overlap rate of the pupil images. First of all, one can note that as soon as $2f\theta/D>1.5$, the 3 curves ($\textcolor{red}{\textbf{s}}$, $\textbf{d}$ and $\textcolor{blue}{\textbf{s.d}}$) do not evolve anymore and this is a general behavior observed for all the configuration tested in this article. In other words, $\theta$ has an influence only when the pupil images overlap.

Secondly, as mentionned in letter \cite{Fauvarque2015}, figure \ref{SD_pyr} shows that an optical recombination induced by a small angle provides a better sensitivity in the high spatial frequencies while it decreases for the low frequencies. Moreover, it is possible to choose where the sensitivity is maximum by changing the $\theta$ value. It comes as no surprise that the linearity range has an inverse behavior: it improves for the low frequencies and decreases at the high ones. The curve of the SD factor is more interesting since for small angles, this curve stays above the classical pyramid one for all the spatial frequencies. In terms of trade-off between sensitivity and linearity range, there is thus a real gain to using small angles.   

\subsubsection{Modulation radius}

\begin{figure}[htb]
\centering
\rotatebox{90}{\hspace{2.5cm}\textcolor{red}{\textbf{s}}~~~~ \textcolor{blue}{\textbf{s.d}}~~~~\textbf{d} }
\includegraphics[trim = 1cm 1.5cm 1.7cm 1.8cm, clip,width=8cm]{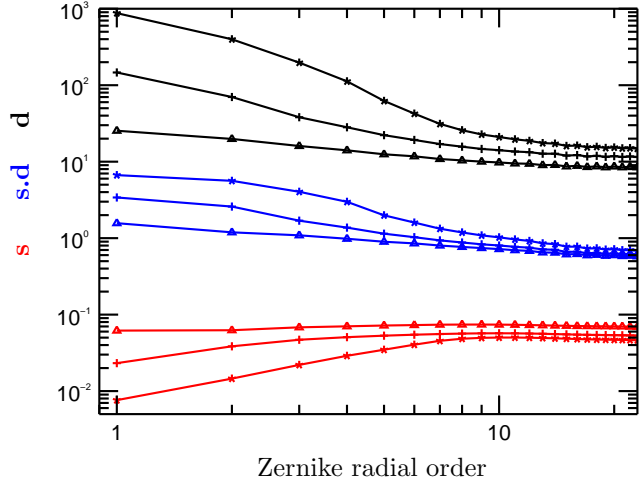}\\
\hspace{1.25cm}Zernike radial order
\caption{\textcolor{red}{\textbf{Sensitivity}}, \textbf{Linearity range} and \textcolor{blue}{\textbf{SD factor}} of the modulated Pyramid WFS with respects to the spatial frequencies. \textbf{Modulation radius} equals to 0 ($\triangle$), 1 ($+$) and 3 ($*$) $\lambda/D$.  Apex angle equals $3\frac{2\theta f}{D}$, i.e. the 4 pupil images are widely separated. \label{SD_pyr_mod}}
\end{figure}

The Pyramid WFS is usually coupled with a modulation stage. With the formalism introduced in section \ref{ttt}, the classical modulation used with Ragazzoni's pyramid, i.e. a \textbf{circular and uniform tip/tilt modulation} with an modulation radius equals to $r_m$, corresponds to:
\begin{equation*}
\phi_m(s) = r_m \left[\cos(2\pi s) Z_1^{-1} + \sin(2\pi s) Z_1^{1}\right] \text{~~~and~~~} w(s) = 1
\end{equation*}
We consider here that the apex angle $\theta$ equals to $3\frac{2\theta f}{D}$, the pupil images are thus widely separated (right insert of figure  \ref{Fc_ecart}). Figure \ref{SD_pyr_mod} shows the sensitivity, the linearity range and the SD factor  with respect to the first 24 first Zernike radial orders for different modulation radius. 
One can observe the classical influence of the modulation radius on the sensitivity and the linearity range: there is a loss a sensitivity for the low spatial frequencies  with a slope sensor behavior on this range. The cut-off frequency is growing linearly with the modulation radius. The linearity range is also improving with the modulation radius. We observe thus that our definition of the sensitivity and the linearity range allows to get, in an analytic way, the usual behaviors of the modulated Pyramid WFS.
By looking at the SD factor which characterizes the trade-off sensitivity/linearity range, it appears that the gain is particularly pronounced for the low spatial frequencies but tends to be null for the highest ones.

\subsection{Mathematical developments applied on the Ragazonni's PWFS }

The aim of this paragraph is to apply the analytical developments to the non-modulated Pyramid WFS. In particular, we show how our formalism may be linked to the classical meta-intensities used for the PWFS  which are usually called \textbf{"Slopes maps"} \cite{Ragazzoni1996}, \cite{Verinaud2004}.

We assume that the apex angle tends to infinity. As a consequence, the 4 pupil images are \emph{completely} separated and do not interfere. Such a theoretical framework allows to study the 4 quadrants independently, subsequently each pupil image has its own intensity: 
\vspace*{-0.15cm}
\begin{eqnarray*}
I^{-+}=|\psi_p \star \mathcal{F}[\mathbb{I}_{\Omega^{-+}}]|^2~~~~~~~~~~I^{++}=|\psi_p \star \mathcal{F}[\mathbb{I}_{\Omega^{++}}]|^2\\
I^{--}=|\psi_p \star \mathcal{F}[\mathbb{I}_{\Omega^{--}}]|^2~~~~~~~~~~I^{+-}=|\psi_p \star \mathcal{F}[\mathbb{I}_{\Omega^{+-}}]|^2
\end{eqnarray*}
The first step consists in getting the linear and quadratic intensities associated to the general definition of the meta-intensities (see equation \ref{MI}).
Thanks to the equation \ref{mIl}, it is possible to determine the phase \emph{linear} dependence of the meta-intensity associated to one of the 4 pupil images. Let's take the case of the pupil image containing the $x$-positive and $y$-positive spatial frequencies, i.e. the $\Omega^{++}$ quadrant. The linear intensity is:
\begin{equation*}
I_l^{++}(\phi_t) =  2\Im[(\mathbb{I}_P \star CIR[\mathbb{I}_{\Omega^{++}}])(\overline{\mathbb{I}_P \phi_t \star CIR[\mathbb{I}_{\Omega^{++}}]})]
\end{equation*}
The development of this equation gives: 
\begin{equation*}
I_l^{++}(\phi_t)= \frac{1}{8}\left[\left(\mathcal{I}+\mathcal{H}^2_{xy}\right)[\mathbb{I}_P]\left(\mathcal{H}_x+\mathcal{H}_{y}\right)[\mathbb{I}_P\phi_t]\right]-\frac{1}{8}\left[\left(\mathcal{I}+\mathcal{H}^2_{xy}\right)[\mathbb{I}_P \phi_t]\left(\mathcal{H}_x+\mathcal{H}_{y}\right)[\mathbb{I}_P]\right]
\end{equation*}
where the operators $\mathcal{I}$, $\mathcal{H}_x$, $\mathcal{H}_y$ and $\mathcal{H}_{xy}^2$ are defined in the appendix \ref{notations}. 
The linear intensities associated to the other quadrants are:
\begin{equation*}
I_l^{+-}(\phi_t)= \frac{1}{8}\left[\left(\mathcal{I}-\mathcal{H}^2_{xy}\right)[\mathbb{I}_P]\left(\mathcal{H}_x-\mathcal{H}_{y}\right)[\mathbb{I}_P\phi_t]\right]-\\
\frac{1}{8}\left[\left(\mathcal{I}-\mathcal{H}^2_{xy}\right)[\mathbb{I}_P \phi_t]\left(\mathcal{H}_x-\mathcal{H}_{y}\right)[\mathbb{I}_P]\right]
\end{equation*}
\begin{equation*}
I_l^{-+}(\phi_t)= \frac{1}{8}\left[\left(\mathcal{I}-\mathcal{H}^2_{xy}\right)[\mathbb{I}_P]\left(-\mathcal{H}_x+\mathcal{H}_{y}\right)[\mathbb{I}_P\phi_t]\right]-\\ \frac{1}{8}\left[\left(\mathcal{I}-\mathcal{H}^2_{xy}\right)[\mathbb{I}_P \phi_t]\left(-\mathcal{H}_x+\mathcal{H}_{y}\right)[\mathbb{I}_P]\right]
\end{equation*}
\begin{equation*}
I_l^{--}(\phi_t)= \frac{1}{8}\left[\left(\mathcal{I}+\mathcal{H}^2_{xy}\right)[\mathbb{I}_P]\left(-\mathcal{H}_x-\mathcal{H}_{y}\right)[\mathbb{I}_P\phi_t]\right]-\\
\frac{1}{8}\left[\left(\mathcal{I}+\mathcal{H}^2_{xy}\right)[\mathbb{I}_P \phi_t]\left(-\mathcal{H}_x-\mathcal{H}_{y}\right)[\mathbb{I}_P]\right]
\end{equation*}
These equations allow to have an analytic expression of the sensibility associated to the turbulent phase mode $\phi_t$:
\begin{equation*}
s(\phi_t)=||I_l^{+-}(\phi_t)||_2+||I_l^{-+}(\phi_t)||_2+\\||I_l^{--}(\phi_t)||_2+||I_l^{++}(\phi_t)||_2
\end{equation*}
Concerning the quadratic intensities range, we choose to not explicitly show their expressions due to the fact that they may seem "abstruse" but, obviously, it is  possible to get them thanks to equation \ref{mIq}. We deduce from them the linearity range: 
\begin{equation*}
d(\phi_t)=\Big(||I_q^{+-}(\phi_t)||_2+||I_q^{-+}(\phi_t)||_2+||I_q^{--}(\phi_t)||_2+||I_q^{++}(\phi_t)||_2\Big)^{-1}
\end{equation*}

\begin{figure}[htb]
\centering
\rotatebox{90}{\hspace{0.75cm}$G_{el}^S$ and $G_{el}^{mI}$~~~ Gap Effective/Linear }
\includegraphics[trim = 1cm 1.5cm 1.7cm 2cm, clip,width=8cm]{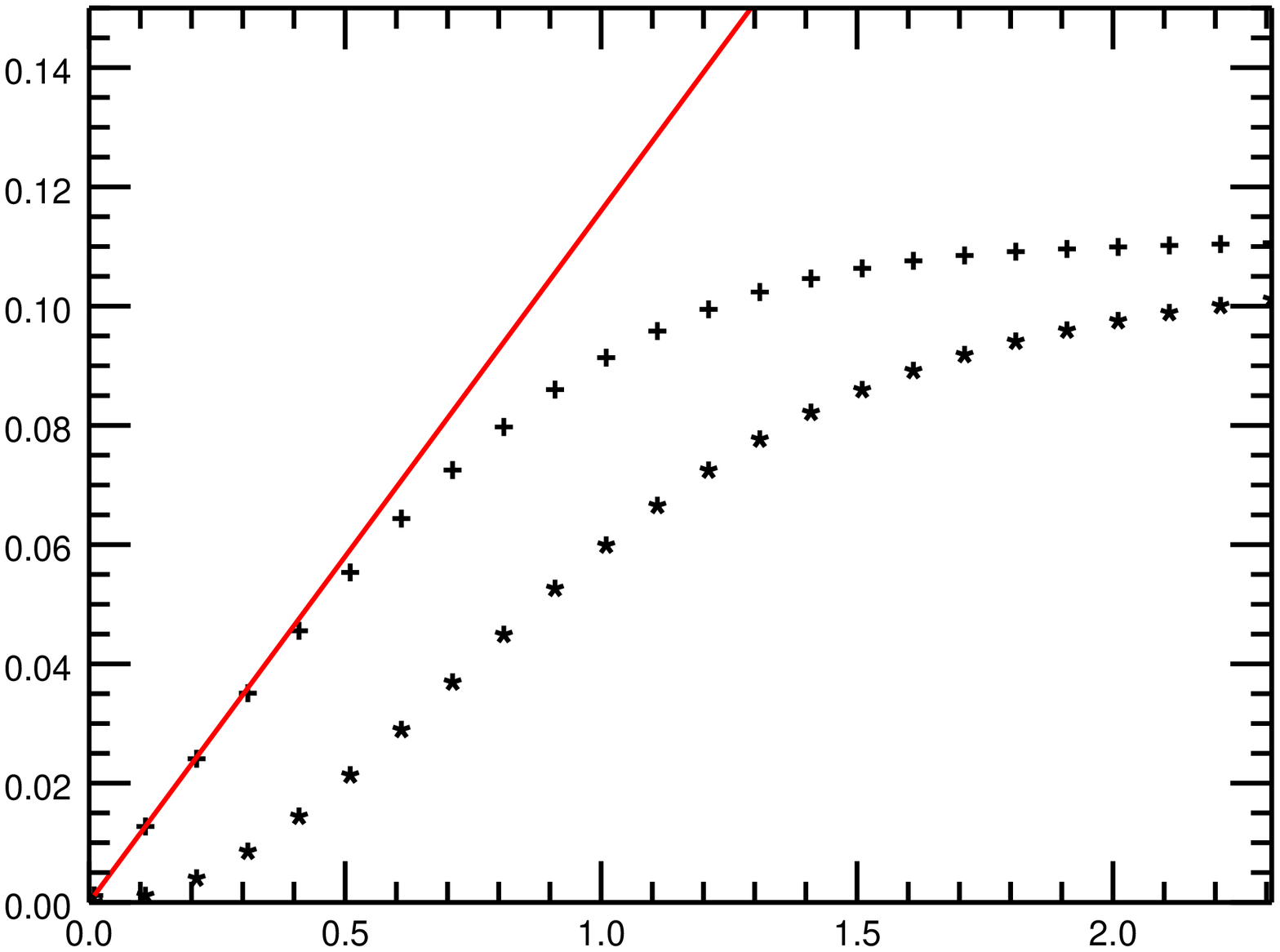}\\
$a$~~~ Input phase amplitude (radian RMS)
\caption{Distance between the effective and linear meta-intensities ($+$, $G_{el}^{mI}$) and the effective and linear slopes maps ($*$, $G_{el}^S$) in function of the input phase amplitude. The slope of the red line equals to the 2-norm of the quadratic intensity associated to the meta-intensities. The input phase is the vertical coma and the considered WFS is the PWFS with an infinite apex angle.\label{GAP_le_pwfs}}
\end{figure}

The general meta-intensities we introduced above (equation \ref{MI}) do not correspond to the classical processing made on intensity in order to create a quantity linear with the incoming turbulent phase. Indeed, it is the custom to use the \textbf{slopes maps} $S^x$ and $S^y$. These quantities are directly calculated from the 4 intensities on the detector. There are two ways to define them (Ragazzoni \cite{Ragazzoni1996} and Verinaud \cite{Verinaud2004}) depending on the normalization but we are interested here in the Verinaud's one:
\begin{equation}
S^x = \frac{I^{++}+I^{+-}-I^{-+}-I^{--}}{n}  \label{Slopesx}
\end{equation}
\begin{equation}
S^y = \frac{I^{++}-I^{-+}+I^{+-}-I^{--}}{n} \label{Slopesy}
\end{equation}
One can note that this definition does not require a return-to-reference. However it is only valid if the reference phase is the null phase. In that particular case,  the slopes maps are related to our meta-intensities via the \textbf{linear} transformations:
\begin{equation}
S^x = mI^{++}+mI^{+-}-mI^{-+}-mI^{--} \label{mat_orthox}
\end{equation}
\begin{equation}
S^y = mI^{++}-mI^{-+}+mI^{+-}-mI^{--}\label{mat_orthoy}
\end{equation}
Subsequently, it is possible to get the \emph{linear} and \emph{quadratic} slopes maps:
\begin{equation*}
S^x_l = I_l^{++}+I_l^{+-}-I_l^{-+}-I_l^{--}~~~~~~~S^y_l = I_l^{++}-I_l^{-+}+I_l^{+-}-I_l^{--}
\end{equation*}
\begin{equation*}
S^x_q = I_q^{++}+I_q^{+-}-I_q^{-+}-I_q^{--}~~~~~~~S^y_q = I_q^{++}-I_q^{-+}+I_q^{+-}-I_q^{--}
\end{equation*}
It is thus possible to have the analytic expression of the linear slopes maps:
\begin{equation*}
S^x_l(\phi_t) = \frac{1}{2}\left(\mathcal{I}[\mathbb{I}_P]\mathcal{H}_x[\mathbb{I}_P\phi_t]+\mathcal{H}_{xy}^2[\mathbb{I}_P]\mathcal{H}_y[\mathbb{I}_P\phi_t] \right)-\\
\frac{1}{2}\left(\mathcal{I}[\mathbb{I}_P\phi_t]\mathcal{H}_x[\mathbb{I}_P]-\mathcal{H}_{xy}^2[\mathbb{I}_P\phi_t]\mathcal{H}_y[\mathbb{I}_P]\right)
\end{equation*}
\begin{equation*}
S^y_l(\phi_t) = \frac{1}{2}\left(\mathcal{I}[\mathbb{I}_P]\mathcal{H}_y[\mathbb{I}_P\phi_t]+\mathcal{H}_{xy}^2[\mathbb{I}_P]\mathcal{H}_x[\mathbb{I}_P\phi_t] \right)-\\
\frac{1}{2}\left(\mathcal{I}[\mathbb{I}_P\phi_t]\mathcal{H}_y[\mathbb{I}_P]-\mathcal{H}_{xy}^2[\mathbb{I}_P\phi_t]\mathcal{H}_x[\mathbb{I}_P]\right)
\end{equation*}
These results already appeared in Shatokhina et al. \cite{Shat11}. They correspond to the 2D generalization of the Verinaud's calculation of article \cite{Verinaud2004} and allow to derive the sensitivity associated to the slopes maps that we call $s_{S}(\phi_t)$:
\begin{equation*}
s_{S}(\phi_t) = ||S_l^x(\phi_t)||_2+||S_l^y(\phi_t)||_2
\end{equation*}
Very fortunately, the loss of sensitivity due to the computation of the slopes maps is insignificant and the equality $s_S(\phi_t)= s(\phi_t)$ may be considered as true. This is surprising since a part of information is lost during the transformations \ref{mat_orthox} and \ref{mat_orthoy}. This remarkable fact will be discussed in an article dedicated to the numerical handling of the meta-intensities (see Fauvarque et al. \cite{Fauvarque2016prep} in preparation).\\ 
On the other hand, the calculation of the quadratic slopes maps show that they equal to zero:
\begin{equation*}
S^x_q(\phi_t) = 0 \text{~~~~and~~~~} S^y_q(\phi_t) = 0
\end{equation*}
The linear transformations \ref{mat_orthox} and \ref{mat_orthoy} have thus a positive influence on the linearity range without any significant loss of sensitivity ! This result constitutes a strong argument in favour of the slopes maps. Unfortunately, it does not mean that the PWFS associated to the slopes maps has an infinite linearity range: if the quadratic intensity is null, it is not the case for the next phase powers. In other words, the linearity range associated to the slopes maps are determined by the cubic intensity, i.e. the third term ($q=3$) of equation \ref{CauchyLaww}. Such a fact is visible on figure \ref{GAP_le_pwfs}: the slope at the origin of the distance between the effective and linear slopes maps which is defined in equation \ref{Gels} is null but this distance does not equal to zero.
\begin{equation}
G_{el}^S(\phi_t,a) = \left|\left|\frac{S_x(a\phi_t)}{a}- S_l^x(\phi_t)\right|\right|_2+\left|\left|\frac{S_y(a\phi_t)}{a}- S_l^y(\phi_t)\right|\right|_2 \label{Gels}
\end{equation}
\begin{equation}
G_{el}^{mI}(\phi_t,a) =\sum_{\text{4 pupils}}~\left|\left|\frac{mI^ {\pm\pm}(a\phi_t)}{a}- I_l^{\pm\pm}(\phi_t)\right|\right|_2 \label{GelmI}
\end{equation}

\section{Application to the Zernike WFS}

\subsection{Tessellation formalism}

This section focus on the Zernike WFS initially introduced  by Zernike himself \cite{zer1934}. This WFS allows us to illustrate the piston presence in the tessellation formalism. Indeed this sensor is working by splitting the incoming energy into two parts thanks to a polar tessellation $\Omega_\rho$ and $\bar{\Omega}_\rho$ and by creating an optical path difference between these two contributions. Note that there is no rejection tip/tilt angles for the Zernike WFS. Hence, we can write:
\begin{equation*}
\Omega_\rho :~~\delta, 0, 0 ~~~~~~~~~~~~~~~~~~\bar{\Omega}_\rho :~~0, 0, 0
\end{equation*}
As a consequence, the 2DFT of the Zernike mask with the optical parameters $\delta$ and $\rho$ is:
\begin{eqnarray}
\mathcal{F}[m_Z](r_d) = &\mathcal{F}[\mathbb{I}_{\bar{\Omega}_\rho}](r_d)+\exp\left(\frac{2\imath\pi}{\lambda_0} \delta \right)\mathcal{F}[\mathbb{I}_{\Omega_\rho}](r_d) \\
&= \delta(r_d) + (\exp\left(\frac{2 \imath \pi}{\lambda_0} \delta\right)-1)\frac{\rho}{r_d} J_1(2\pi \rho r_d) \label{zer_CIR}
\end{eqnarray}
Historically, the size of the central mask $\rho$ was calculated in order to have the same amount of energy in $\Omega_\rho$ and $\bar{\Omega}_\rho$, i.e. $\rho=1.06\lambda_0/D$ for a circular pupil; 
the depth $\delta$ associated to the piston was set to create a $\pi/2$ phase gap between the two waves coming from $\Omega_\rho$ and $\bar{\Omega}_\rho$.
Nevertheless, the previous set of optical parameters is not the only one which generates a WFS. $\rho$ provides the energetic partition between the two parts of the polar tessellation and an exact equality is actually not strictly needed: a certain leeway exits and allows to modify some properties of the sensor. In the same way, recent works led by N'diaye et al. \cite{Diaye2014} shows that $\delta$ may range on the interval $[-\lambda_0/8; 3\lambda_0/8]$ and still defines a WFS. Consequently, as for the Pyramid WFS, the historical Zernike sensor is only a member of a vaster \emph{class} of sensors generated by the polar tessellation and the two optical parameters $\rho$ and~$\delta$.

By looking at equation \ref{zer_CIR}, it appears that the entrance pupil is convoluted with the $J_1$ Bessel function. As a consequence, we define the following operator that we call the Zernike operator~$\mathcal{Z}_\rho$:
\begin{equation*}
\mathcal{Z}_{\rho}[f] = f \star \frac{\rho}{r_d} J_1(2\pi \rho r_d) 
\end{equation*}
Thanks to the general definition of the linear intensity $I_l$ (see equation \ref{mIl}), it is possible to get the purely linear behavior of the ZWFS:
\begin{equation}
I_l(\phi_t) = \overbrace{2\sin\left(\frac{2\pi}{\lambda_0}\delta\right)}^a\overbrace{\mathbb{I}_P \Big(\phi_t\mathcal{Z}_{\rho}[\mathbb{I}_P]-\mathcal{Z}_{\rho}[\mathbb{I}_P\phi_t]  \Big)}^b \label{CCIRZ}
\end{equation} 
The spatial variability of the linear intensity $I_l$ is coded by the term $b$ of equation \ref{CCIRZ}. One notes that this variability only depends on the optical parameter $\rho$ which is the size of the Zernike mask. The depth of this mask coded by the optical parameter $\delta$ allows to adjust the global scalar factor $a$. 


The quadratic intensity $I_q$ is calculated via its definition in equation \ref{mIq}:
\begin{equation*}
I_q(\phi_t) = \left[1-\cos\left(\frac{2\pi}{\lambda_0}\delta\right)\right]\Big(2\mathcal{Z}_\rho^2[\phi_t]-
2\phi_t\mathcal{Z}_\rho[\phi_t]+\\
\mathcal{Z}_\rho[\phi_t^2]+\phi_t^2\mathcal{Z}_\rho[\mathbb{I}_P]-2\mathcal{Z}_\rho[\mathbb{I}_P]
\mathcal{Z}_\rho[\phi_t^2]\Big)
\end{equation*}
Once again, it appears that $\rho$ only influences the spatial variability of the quadratic intensity whereas $\delta$ adjusts a global scalar factor. 

\subsection{Sensitivity, linearity range and SD factor}

From the expression of the linear intensity, we can get the sensitivity associated to the input phase $\phi_t$:
\begin{equation}
s(\phi_t) = 2\left|\sin\left(\frac{2\pi}{\lambda_0}\delta\right)\right| || \phi_t\mathcal{Z}_{\rho}[\mathbb{I}_P]-\mathcal{Z}_{\rho}[\mathbb{I}_P\phi_t]   ||_2\label{SenZernike}
\end{equation}
The linearity range is directly related to the inverse of the 2-norm of the quadratic intensity:
\begin{equation}
d(\phi_t) =\frac{1}{1-\cos\left(\frac{2\pi}{\lambda_0}\delta\right)} ||2\mathcal{Z}_\rho^2[\phi_t]-2\phi_t\mathcal{Z}_\rho[\phi_t]+\\
\mathcal{Z}_\rho[\phi_t^2]+\phi_t^2\mathcal{Z}_\rho[\mathbb{I}_P]-2\mathcal{Z}_\rho[\mathbb{I}_P]
\mathcal{Z}_\rho[\phi_t^2]||_2^{-1} \label{dynZernike}
\end{equation}

\subsubsection{Depth of the Zernike mask}
\begin{figure}[htb]
\centering
\rotatebox{90}{\hspace{2.5cm}\textcolor{red}{\textbf{s}}~~~~ \textcolor{blue}{\textbf{s.d}}~~~~\textbf{d} }
\includegraphics[trim = 1cm 1.5cm 1.7cm 1.8cm, clip,width=8cm]{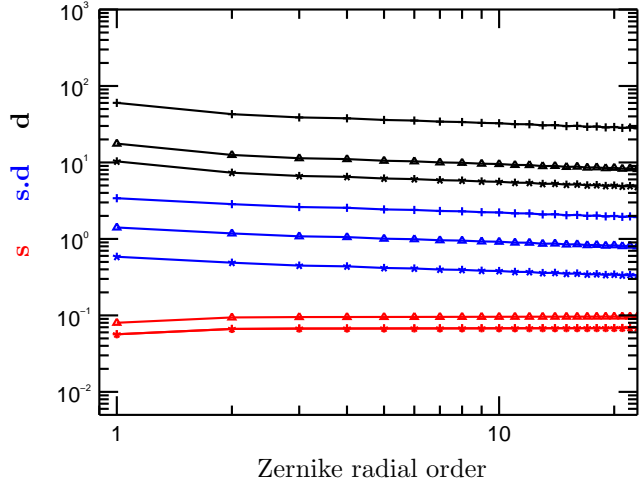}\\
\hspace{1.25cm}Zernike radial order
\caption{\textcolor{red}{\textbf{Sensitivity}}, \textbf{Linearity range} and \textcolor{blue}{\textbf{SD factor}} of the Zernike WFS with respects to the spatial frequencies. \textbf{Depth of the Zernike mask $\delta$} equals to 1/8 ($+$), 1/4 ($\triangle$) and 3/8 ($*$) $\lambda_0$. Sensitivity is identical for $\delta$ equals to 1/8 and 3/8 $\lambda_0$. \label{SD_Z_dp}}
\end{figure}

The size of the Zernike mask $\rho$ is set at $1.06\lambda_0/D$. We study the influence of its depth on the sensitivity, the linearity range and the SD factor.  
Equation \ref{SenZernike} shows that the sensibility is maximum for $\delta=\lambda_0/4$. That corresponds to the historical Zernike WFS.  From $\delta=0$ to $\delta=\lambda_0/2$, we then observe thanks to equation \ref{dynZernike} that the linearity range is decreasing. 
When $\delta$ tends to 0, the linearity range is infinite but such a configuration corresponds to the trivial mask, the sensibility is then null. Figure \ref{SD_Z_dp} confirms these results. Moreover, the SD factor curve shows that deep Zernike masks, e.g. $3/8\lambda_0$, allow to get a significant gain in terms of linearity range without too much sensitivity loss. Finally, we effectively observe that the shape of the curves does not change when $\delta$ varies. This parameter only translates them by the global factors of equations \ref{SenZernike} and \ref{dynZernike}.

\subsubsection{Size of the Zernike mask}
We set in this paragraph the depth of the Zernike mask at $\lambda_0/4$ and study the influence of its size on the performance criteria.
\begin{figure}[htb]
\centering
\rotatebox{90}{\hspace{2.5cm}\textcolor{red}{\textbf{s}}~~~~ \textcolor{blue}{\textbf{s.d}}~~~~\textbf{d} }
\includegraphics[trim = 1cm 1.5cm 1.7cm 1.8cm, clip,width=8cm]{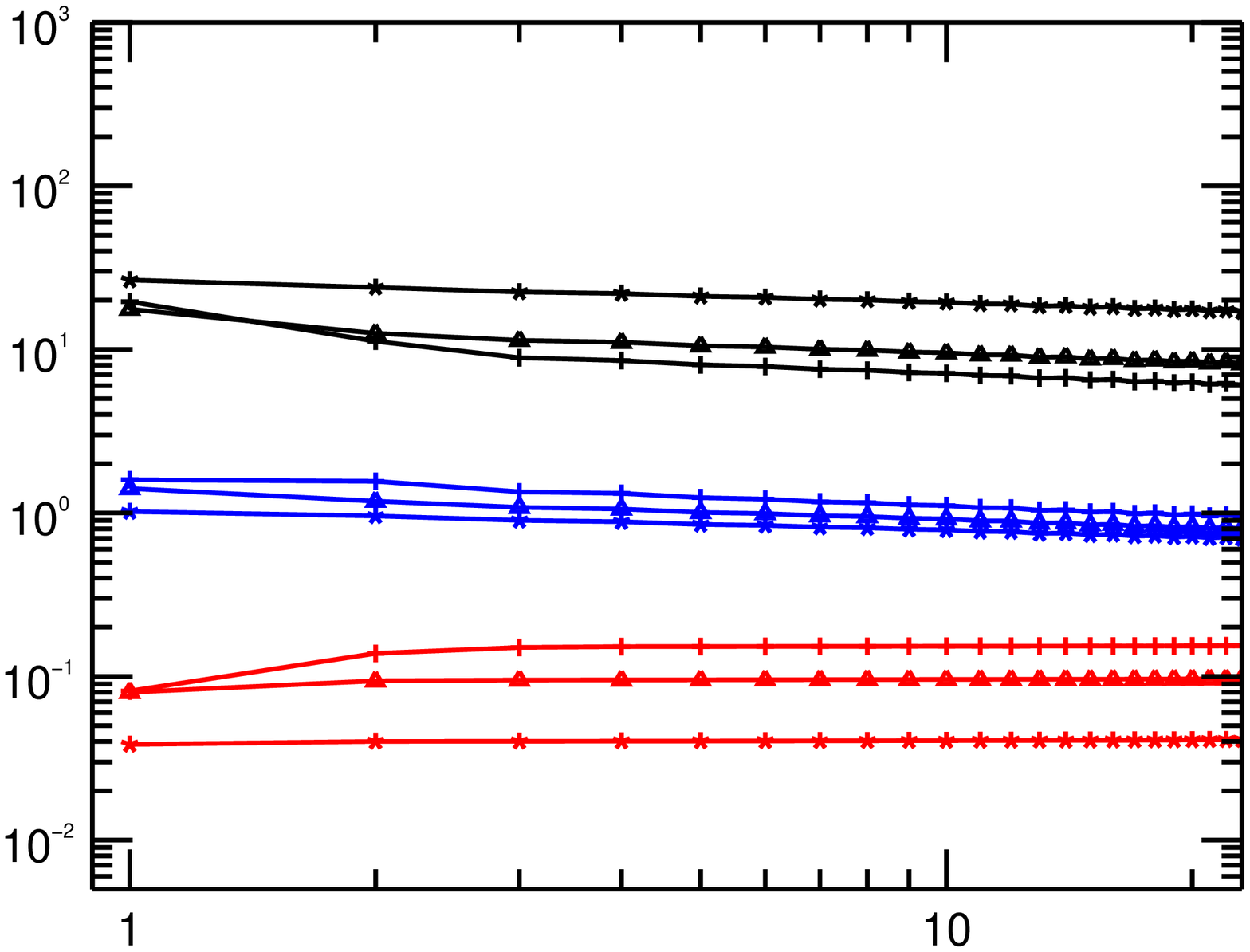}\\
\hspace{1.25cm}Zernike radial order
\caption{\textcolor{red}{\textbf{Sensitivity}}, \textbf{Linearity range} and \textcolor{blue}{\textbf{SD factor}} of the Zernike WFS with respects to the spatial frequencies. \textbf{Size of the Zernike mask $\rho$} equals to 0.5 ($*$), 1 ($\triangle$) and 1.5 ($+$) 1.06$\lambda_0/D$.\label{SD_Z_sd}}
\end{figure}
We observe on figure \ref{SD_Z_sd} that this time, the shape of curves changes with $\rho$. This is fully consistent with equations \ref{SenZernike} and \ref{dynZernike}. Moreover, a larger Zernike mask slightly improves the sensitivity while the linearity range decreases. The SD factor approximately stays constant. 
%
%
We conclude this part about the Zernike WFS class by remarking that the first optical stage of the Roddier\&Roddier coronagraph is also a member of this class. 

\section{Polychromatic light}\label{pooly}

In this section we explore the influence of an incoming polychromatic light on the previous mathematical developments. 

\subsection{Polychromatic intensity}
The first step consists in writing all the previous formula by making explicit the dependence with the wavelength $\lambda$. The incident EM field is:
\begin{equation*}
\psi_p(x_p,y_p,\lambda) = \sqrt{n(\lambda)} \mathbb{I}_P(x_p,y_p) exp \left( \frac{2\imath\pi}{\lambda}\Delta(x_p,y_p) \right)
\end{equation*}
with $n(\lambda)$ defined in such a way that $n(\lambda)d\lambda$ is the number of photons by unit area on the wavelength range $[\lambda-d\lambda/2;\lambda+d\lambda/2]$. In other words, $n(\lambda)$ is the spectrum of the studied source. The optical path difference $\Delta$ is not dependent on the wavelength since we assume that the atmosphere is not a dispersive media. If we choose a particular wavelength $\lambda_0$ as a reference, it is possible to write the phase of the incoming field as:
\begin{equation}
\phi = \frac{2\pi}{\lambda}(\Delta_t+\Delta_r) = \frac{2\pi}{\lambda_0} \frac{\lambda_0}{\lambda}(\Delta_t+\Delta_r) \hat{=}\frac{\lambda_0}{\lambda}( \phi_t+\phi_r) \label{eqq}
\end{equation}
where $\phi_t$ (resp. $\phi_r)$ is the turbulent phase corresponding to the turbulent OPD $\Delta_t$ (resp. reference OPD $\Delta_r$) at the wavelength $\lambda_0$. In other words, $\lambda_0$ is the analysis wavelength. Equation \ref{eqq} also means that the only impact of polychromatism on the incoming turbulent phase corresponds in a scale factoring. Polychromatism does not mix two different turbulent phase modes together. This fact is essential for the continuation of our study.

The transparency function of the mask does not change but we now keep its $\lambda$-dependency:
\begin{equation*}
m(f_x, f_y,\lambda) = exp\left(\frac{2\imath\pi}{\lambda}OS(f_x,f_y,\lambda)\right)
\end{equation*}
The Optical Shape of the mask depends, a priori, on the wavelength since this quantity is the product of the refractive index $n_r$ (which depends on $\lambda$ for dispersive material) and the Geometrical Shape of the mask $GS$:
\begin{equation*}
OS(f_x,f_y,\lambda)=n_r(\lambda)GS(f_x,f_y)
\end{equation*} 
The last source of $\lambda$-dependency comes from the Fresnel formalism, indeed the light propagation itself includes the wavelength as we can see in the expression of the EM field  in the detector plane:
\begin{equation*}
\psi_d(x_d,y_d) \propto \iint \frac{d x_p dy_p}{f^2} ~~\psi_p(x_p-x_d,y_p-y_d,\lambda)\\
\iint\frac{df_xdf_y}{\lambda^2}~~m(f_x,f_y,\lambda)~\exp \left(-\frac{2\imath\pi}{f\lambda}\left[x_p f_x+y_p f_y\right]\right)\label{nllll}
\end{equation*}
Concretely, it means that the right part integral of this equation, which corresponds to a $\lambda$-dependent 2DFT of the mask. In other words, without particular assumption about the mask, its Fourier transform depends on the wavelength. The fundamental equation \ref{nl} becomes:
\begin{equation*}
I_\lambda(\phi_t) = |\psi_p(\phi_t,\lambda) \star \mathcal{F}_\lambda[m]|^2 
\end{equation*}
Under these new assumptions, we can write the polychromatic intensity on the detector, called $I_p$, as the integral of the monochromatic intensities on the whole spectrum of the light source: 
\begin{equation}
I_p(\phi_t) = \int d\lambda |\psi_p(\phi_t,\lambda) \star \mathcal{F}_\lambda[m]|^2 \label{intensitypoly}
\end{equation}

\subsection{Substitution test}

The chromatic 2DFT equals to:
\begin{equation*}
\mathcal{F}_{\lambda}[m](x_p,y_p) = \iint\frac{df_xdf_y}{\lambda^2}~~m(f_x,f_y,\lambda)~e^{-\frac{2\imath\pi}{f\lambda}\left[x_p f_x+y_p f_y\right]}
\end{equation*}
By looking at this equation, it appears that the chromatic 2DFT may be decoupled from the wavelength on the condition that the substitution
\begin{equation}
(u,v) = \left(\frac{f_x}{\lambda},\frac{f_y}{\lambda}\right)\label{substitution}
\end{equation}
makes disappear the $\lambda$-dependency. This condition requires in fact that the transparency function of the mask is a function of only two variables $u$ and $v$ instead of three: $f_x$, $f_y$ and $\lambda$. 
In terms of physics, such a condition means that the point spread functions for every wavelength always see the same "mask". In other words, the Optical Shape of this mask is invariable by scale change. In particular, one can note that as soon as a mask does not need any \textbf{characteristic length} quantity in its transparency function, the substitution of equation \ref{substitution} is possible.

Under the substitution  of equation \ref{substitution} assumption, the 2DFT becomes independent on the wavelength. The total polychromatic intensity (equation \ref{intensitypoly}) may be simplified into:
\begin{equation*}
I_p(\phi_t) = \int n(\lambda)d\lambda  \left|\mathbb{I}_P \exp\left(\imath\frac{\lambda_0}{\lambda}(\phi_t+\phi_r)\right) \star \mathcal{F}[m]\right|^2
\end{equation*}
The phase power series of such an intensity becomes thus:
\begin{equation}
I_p(\phi_t)=\sum_{q=0}^{\infty}\frac{(-\imath)^q}{q!} \int n(\lambda)\left(\frac{\lambda_0}{\lambda}\right)^q\sum_{k=0}^{q} (-1)^{k} {q \choose k} \phi_t^{k\star} \overline{\phi_t^{q-k\star}}d\lambda~~~~~
\text{where}~~~~~ \phi^{k\star}_t \hat{=} (\mathbb{I}_p ~e^{\frac{2\imath\pi\lambda_0}{\textcolor{red}{\lambda}} \phi_r}~\phi_t^k) \star \mathcal{F}[m]\label{Caupolyy}
\end{equation}
\textbf{We assume, from now, that $\phi_r=0$}. As a consequence, the $k$-th moment does not depend on $\lambda$ anymore. Subsequently, the phase power series \ref{Caupolyy} may be simplified into:
\begin{equation}
I_p(\phi_t)=\sum_{q=0}^{\infty}\frac{(-\imath)^q}{q!} \int n(\lambda)\left(\frac{\lambda_0}{\lambda}\right)^q d\lambda\sum_{k=0}^{q} (-1)^{k} {q \choose k} \phi_t^{k\star} \overline{\phi_t^{q-k\star}}~~~~~
\text{where}~~~~~ \phi^{k\star}_t \hat{=} (\mathbb{I}_p~\phi_t^k) \star \mathcal{F}[m]\label{Caupoly}
\end{equation}

\subsection{Polychromatic meta-intensities}\label{ttet}

The first terms of the previous equation corresponds once again to the constant, linear and quadratic terms, called $I_{pc}$, $I_{pl}$ and $I_{pq}$:
\begin{equation*}
I_{pc}=|\mathbb{I}_P  \star \mathcal{F}[m]|^2
\end{equation*}
\begin{equation*}
I_{pl}(\phi_t)=\frac{1}{n}\left(\int n(\lambda)\frac{\lambda_0}{\lambda}d\lambda\right) 2\Im[(\mathbb{I}_P \star \mathcal{F}[m])(\overline{\mathbb{I}_P~\phi_t \star \mathcal{F}[m]})]
\end{equation*}
\vspace*{-0.5cm}
\begin{equation*}
I_{pq}(\phi_t)=\frac{1}{n}\left(\int n(\lambda)\left(\frac{\lambda_0}{\lambda}\right)^2 d\lambda\right)\Big[|\mathbb{I}_P~ \phi_t \star \mathcal{F}[m]|^2]-\\
\Re[(\mathbb{I}_P~\star \mathcal{F}[m])(\overline{\mathbb{I}_P~\phi^2_t \star \mathcal{F}[m]})]\Big]
\end{equation*}
where $n$ is the total flux:
\begin{equation*}
n= \int n(\lambda) d\lambda
\end{equation*}
The method used previously to construct a phase-linear quantity (at least in the small phases approximation) is once again valid. The effective polychromatic meta-intensity, called $mI_p$, will be subsequently defined in the same way as equations \ref{MI} and \ref{MIm}, i.e.: 
\begin{equation}
mI_p(\phi_t)=\frac{I_p(\phi_t)-I_p(0)}{n}\label{MIp}  
\end{equation}
Subsequently, it is this time easy to do the link between the polychromatic linear and quadratic intensities to the monochromatic ones:
\begin{eqnarray*}
I_{pl}(\phi_t) = &\frac{1}{n}\left(\int n(\lambda)\frac{\lambda_0}{\lambda}d\lambda\right) I_{l}(\phi_t)\\
I_{pq}(\phi_t) = &\frac{1}{n}\left(\int n(\lambda)\left(\frac{\lambda_0}{\lambda}\right)^2d\lambda\right) I_{q}(\phi_t)
\end{eqnarray*}
It appears thus that under the assumption that substitution  of eq. \ref{substitution} is valid and that the reference phase equals to the null phase, the polychromatic linear and quadratic intensity only differs from the \emph{monochromatic} ones by global gain factors which only depends on the spectrum $n(\lambda)$ and on the reference wavelength $\lambda_0$ but not on the incoming phase mode.

In particular, it means that the polychromatism does not induce a change in the spatial structure of the meta-intensities but only introduce a global \emph{blurring factor}. In other words, an interaction matrix done with a monochromatic reference source is still valid for a polychromatic source if such a global factor is taken into account.

Substitution of eq. \ref{substitution} is thus a robust way to define what is an achromatic sensor. Indeed, if this one is not possible the factorization in the power series \ref{Caupolyy} is not possible anymore. Concretely, it means that there is cross talk between the phase modes due to the polychromatism of the source. The interaction matrix will be, subsequently, changed in its spatial structure. 

Finally, one can notice the impact of a non-null reference phase: a WFS which would not work around the null phase is irremediably chromatic.

\subsection{Sensitivity, linear range and SD factor}

We can finally note that, \textbf{for an achromatic sensor}, the polychromatic sensitivity and linearity range (called $s_p$ and $d_p$) may easily be linked to the monochromatic ones:
\begin{equation*}
s_p(\phi_t) = \frac{1}{n} \left(\int n(\lambda)\frac{\lambda_0}{\lambda}d\lambda\right) s(\phi_t)
\end{equation*}
\begin{equation*}
d_p(\phi_t) = n\left(\int n(\lambda)\left(\frac{\lambda_0}{\lambda}\right)^2d\lambda\right)^{-1}d(\phi_t)
\end{equation*}
%

%
%
%

\subsection{Applications to classical WFSs }

As we have just seen in the previous paragraph, as soon as the reference phase equals to zero, the chromatic behavior of a WFS only depends on the expression of the transparency function of its Fourier mask. In this section, we browse the previous examples of WFS and study their chromatic behavior in the light of the substitution test defined in equation \ref{substitution}.

\subsubsection{The Pyramid WFS class}

\paragraph{Perfect pyramidal mask}

We first consider the class of the Pyramid WFSs. The general expression of the transparency function of the associated mask is:
\begin{equation*}
m_\triangle(f_x,f_y,\lambda) = \exp \left(\frac{2\imath\pi}{\lambda} n_r(\lambda)~\theta(|f_x|+|f_y|)\right)
\end{equation*}
There are two cases depending on whether we consider a dispersive Pyramid (e.g. a glass transparent Pyramid) or a reflective Pyramid (Wang et al. \cite{Wang10}). In the first case, $n_r(\lambda)$ follows the usual empirical Cauchy's equation:
\begin{equation*}
n_r(\lambda) = B + \frac{C}{\lambda^2} + ...
\end{equation*}
where $B$ and $C$ depend on the nature of the propagation medium. Note that in this case, it is impossible to make $m$ only a function of the two variables $f_x/\lambda$ and $f_y/\lambda$; it means that the associated WFS is not achromatic. Fortunately, a classical solution to solve this problem exits which consists in using two transparent pyramids attached by their base (Esposito et al. \cite{Espo03}). Such a device allows, in substance, to nullify the first $\lambda$-dependent term $C$ of the refractive index. In this way, the variables substitution becomes possible which makes the WFS achromatic. 
In the case of a reflective pyramid, the propagation medium is the air which is not significantly dispersive on the visible spectrum. Subsequently, the substitution can be made and the simplified transparency function is:
\begin{equation*}
m_\triangle(u,v) = \exp \left(2\imath\pi n_{\text{air}}~\theta(|u|+|v|)\right)
\end{equation*}
These results proves that the reflective Pyramid and the 2 attached Pyramids are achromatic sensors. Moreover, the apex angle $\theta$ stays a free parameter. Subsequently, both the classical PWFS introduced by Ragazonni \cite{Ragazzoni1996} where the pupil images are completely separated and the Flattened PWFS we proposed in \cite{Fauvarque2015} are achromatic sensors.


\paragraph{Modulation and chromaticity}

We already saw (paragraph \ref{ttt}.\ref{refpha}) that the modulated phase $\phi_m(s)$ may be seen as a variable reference phase. Moreover, results of paragraph \ref{pooly}.\ref{ttet} showed that a non-null reference phase irremediably induce a \textbf{structural} dependence of the output meta-intensities regarding to the spectrum of the source. Consequently, the modulated PWFS (and all its variations) is not an achromatic sensor.

\subsubsection{Zernike mask}
The transparency function of the mask of the Zernike WFSs class needs unfortunately 2 characteristic length quantities: the first one is the size of the central disc allowing to define the two parts of the polar tessellation. This size is set in order to set the energy partition in the 2 parts of the tessellation. For a circular pupil of a diameter $D$ and a reference wavelength $\lambda_0$, the size of the pupil equals to 1.06$\lambda_0/D$. The second characteristic quantity is the depth of the central disc well. It is set in order to create an OPD between the two parts of the polar tessellation. In other words, the optical design of the Zernike mask is optimized only for one wavelength. Moreover, this mask is manufactured in a transparent medium which is, a priori, dispersive as well but we do not consider such a chromatic effect for the following study. Such considerations allows us to write the transparency function like:
\begin{equation*}
m_Z(f_x, f_y,\lambda) = 1 + \left(e^{\frac{\imath\pi}{2} \frac{\lambda_0}{\lambda}}-1\right)\Theta\left(1.06\frac{\lambda_0}{D}-\sqrt{f_x^2+f_y^2}\right) 
\end{equation*}
It appears clearly that substitution of eq.  \ref{substitution} does not allow to cancel the wavelength dependence of the mask. Subsequently, the Zernike WFS cannot be an achromatic sensor. Nevertheless, N'Diaye  \cite{Ndia13} showed that chromaticity does not play significant role in terms of error budget making the ZWFS practically achromatic.

\section{Conclusions}

Thanks to the original approach of the Fourier plane tessellation, we managed to unify all the optical designs based on Fourier filtering. 
The essential point is to consider any kind of Fourier mask as a spatial frequencies splitter. 

From this approach, we showed that the 2D Fourier transform of the mask plays a significant role all along the mathematical developments, allowing in particular to efficiently describe the imaging for any kind of entrance pupils or incoming phases. 
A phase power series development of the intensity on the detector allows to identify the linear term.
In order to extract this one from the effective signal obtained on the detector, we introduced a numerical processing which may be seen as the simplest way to get a phase-linear response in the small phases regime. The quantity obtained after such a calculation have been called meta-intensity $mI$ and is defined in the same way for any kind of Fourier based Wave Front sensors.

These analytical developments have then allowed to define the sensitivity and the linearity range of all the WFSs studied here. They are directly based on the linear and quadratic terms of the phase power series of the intensity. 
Moreover, we defined the SD factor which quantifies the trade-off between these two antagonist performance criteria. Such a factor gave us some practical ideas to optimize the Fourier masks depending on requirement specifications. In other words, it opens the way to the free-form filtering masks. 

We then extended all these results to the modulation. This is particularly useful while considering unusual modulation as for example non-uniform or non-tip/tilt modulations.   
 
The last theoretical result was about the effect of a polychromatic incoming light on the efficiency of a WFS. 
We gave a robust criterion, depending only on the transparency function of the mask in order to characterize the achromatic property of a WFS. \\

As a second goal for this paper we applied the theoretical developments to existing and new sensors. Indeed, the developed formalism allowed to unify -- and it was its main purpose -- all the Fourier based WFSs. We showed in particular that the Pyramid WFS and the Zernike WFS may be considered not as unique designs but more as classes, where optical characteristics become flexible parameters. 

It appeared for instance that the Flattened PWFS and the classical PWFS were only two types of the Pyramid WFSs class where the apex angle is considered as a parameter. An exhaustive study which explores in great details all the parameters of the Pyramid class, i.e. the apex angle, the number of faces and the modulation parameters, is led in Fauvarque et al. \cite{FauvarqueSPIE}.

 
We also showed that the slopes maps usually associated to the PWFS were directly linked to the general meta-intensities that we introduced. Moreover, we showed that the linearity range was improved without significant loss of sensitivity thanks to this slopes maps computation. 

Finally, new WFSs have been studied by considering the depth and the size of the central well of the Zernike WFS as a free parameter. Moreover, we isolated, analytically, the role played by each of these two parameters on the response of the sensors of the ZWFS class.


\section*{Funding Information}
This work was co-funded by the European Commission under FP7
Grant Agreement No. 312430 Optical Infrared Coordination Network for Astronomy, by the ANR project WASABI and by the French Aerospace Lab (ONERA) (in the framework of the NAIADE Research Project).

\appendix

\section{Notations}\label{notations}
\begin{eqnarray*}
\text{2D Fourier Transform:}&\hspace{-0.3cm}\mathcal{F}[f](\mu,\nu)=\iint f(x,y) e^{-2\imath\pi (x\mu + y\nu)} dxdy\\
%
%
%
\text{Identity transform:}&\mathcal{I}[f] (x,y) = f(x,y)\\
\text{Symmetric transform:}&\mathcal{S}[f] (x,y) = f(-x,-y)
\end{eqnarray*}
\vspace*{-0.6cm}
\begin{eqnarray*}
&\text{Hilbert transforms:}~~~\mathcal{H}_x[f] (x,y) = p.v.\left\{\frac{1}{\pi} \int \frac{f(t,y)}{x-t} dt\right\}\\
&~~~~~~~~~~~~~~~~~~~~~~~~~~~~~~~~~~~~~\mathcal{H}_y[f] (x,y)  = p.v.\left\{\frac{1}{\pi} \int \frac{f(x,t)}{y-t} dt \right\}\\
&~~~~~~~~~~~~~~~~~~~~~~~~~~~~\mathcal{H}^2_{xy}[f] (x,y)  = p.v.\left\{\frac{1}{\pi^2} \iint \frac{f(t,t')}{(x-t)(y-t')} dtdt' \right\}\\
&\text{2-Norm:} ~~~~~||f||_2 =\left( \iint |f(x,y)|^2 dx dy\right)^{1/2} 
\end{eqnarray*}
\bibliography{library} 
\bibliographystyle{spiebib} 

\end{document}